\UseRawInputEncoding
\pdfoutput=1 \documentclass[aps,prd,amsmath,floats,floatfix, twocolumn,superscriptaddress,nofootinbib,showpacs,longbibliography]{revtex4-2}

\usepackage{amsmath}
\usepackage[T1]{fontenc}
\usepackage[utf8]{inputenc}
\usepackage{lmodern}
\usepackage{cancel}
\usepackage{verbatim}

\usepackage[dvipsnames, usenames]{xcolor}
\definecolor{linkcolor}{rgb}{0.0,0.3,0.5}
\usepackage[hypertexnames=false, unicode, colorlinks=true, linkcolor=linkcolor,citecolor=linkcolor, filecolor=linkcolor, urlcolor=linkcolor, pdfusetitle]{hyperref}

\usepackage[all]{hypcap}
\usepackage{graphicx}
\usepackage{xspace}
\usepackage{amssymb}
\usepackage[normalem]{ulem} 
\usepackage{bm}
\usepackage{float}
\usepackage{placeins}
\usepackage[T1]{fontenc}
\usepackage[utf8]{inputenc}

\usepackage[caption=false]{subfig}

\usepackage{microtype}

\usepackage[english]{babel}
\usepackage{blindtext}

\usepackage{etoolbox} 
\usepackage{orcidlink}

\graphicspath{%
  {figs/}%
}

\DeclareMathAlphabet{\mathpzc}{OT1}{pzc}{m}{it}

\usepackage{soul}

\begin{document}

\preprint{APS/123-QED}

\title{
Constraining $f(R)$ gravity and evolving dark energy via large-scale structure and phase-space trajectories}

\newcommand{\UCT}{\affiliation{Cosmology and Gravity Group, Department of Mathematics and Applied Mathematics, University of Cape Town, Rondebosch 7700, Cape Town, South Africa}}

\newcommand{\USAL}{\affiliation{Departamento de Física Fundamental, Universidad de Salamanca, 37008 Salamanca, Spain}}
\newcommand{\IFT}{\affiliation{Instituto de F\'isica Te\'orica UAM-CSIC, Universidad Aut\'onoma de Madrid, Cantoblanco, 28049 Madrid, Spain}}

\author{Tshepo Mathibela}
\email{mthtsh098@myuct.ac.za}
\UCT 

\author{\'{A}lvaro de la Cruz-Dombriz}
\USAL
\UCT

\author{Savvas Nesseris}
\IFT

\date{\today}

\begin{abstract}
We present a joint observational analysis confronting viable $f(R)$ modified gravity theories, specifically the Hu \& Sawicki and Starobinsky models, with background and large-scale structure (LSS) data. Utilizing Monte-Carlo Markov chain (MCMC) sampling across datasets including baryon acoustic oscillations (BAO), type Ia supernovae (SNeIa), cosmic microwave background (CMB) distance priors, and linear growth measurements ($f\sigma_8$, $f$, $\sigma_8$), we place tight constraints on the model parameters governing deviations from General Relativity. For the full dataset combination, we obtain $\log_{10} b_\mathrm{HS} = -6.325_{-1.138}^{+1.216}$ for the Hu \& Sawicki model and $b_\mathrm{S} = (0.8\pm61.0)\times10^{-4}$ for the Starobinsky model. Model comparison based on the Akaike Information Criterion indicates that these $f(R)$ extensions are statistically favored over flat $\Lambda\text{CDM}$ ($|\Delta\text{AIC}| \ge 3.99$) for the combined data. However, when considering the Bayesian Information Criterion, the evidence for support is significantly reduced. Furthermore, we construct two-dimensional phase-space diagrams in the $(\mu, \gamma)$ and $(\mu, \Sigma)$ planes across several redshifts, establishing a novel diagnostic null-test allowing us to probe for deviations from $\Lambda\text{CDM}$, corresponding to the fixed point $(1,1)$ in both planes, using LSS observables. Should future weak-lensing and galaxy surveys provide data points with $\mu-1<0$ and $\gamma-1>1$ or $\Sigma-1 < 0 $, then the aforementioned models could be directly ruled out. 

\end{abstract}

\maketitle


\section{\label{sec:level1}Introduction}

The Cosmological Concordance $\Lambda$CDM model, based on a cosmological constant $\Lambda$ and cold dark matter (CDM) within a geometric structure as provided by General Relativity (GR) is the most well accepted model for the description of the Universe. Within good precision, it reproduces the observational results that include large-scale structure (LSS) formation \cite{Lopez:2024kih}, the oscillatory nature of baryonic acoustic oscillations (BAO) \cite{DESI:2025zgx}, cosmic microwave background radiation (CMBR) anisotropies \cite{Planck:2018vyg}, and the dimming of type Ia Supernovae (SNeIa) \cite{DES:2024jxu}. However, despite the model's successful description of those cosmological results, it suffers from major fine-tuning \cite{Padmanabhan:2002ji} and, as such, major efforts have been dedicated to the development of other theoretical schemes that can explain the late-time expansion history of the Universe \cite{Abdelwahab:2011dk}.

In the last two decades, modified gravity theories have been suggested as alternatives to the $\Lambda$CDM paradigm. The most popular of those are the $f(R)$ gravity theories \cite{DeFelice:2010aj} which extends the GR Lagrangian by introducing an arbitrary function of Ricci scalar $R$. Those models are known to introduce one extra scalar degree of freedom, dubbed the scalaron, which mediates the fifth force. However, this extra force is severely suppressed by the so-called Chameleon screening mechanism \cite{Khoury:2003aq}. As such, the $f(R)$ gravity models can still explain the late-time cosmic accelerated expansion while satisfying the local gravity tests and the equivalence principle in the highly dense regimes $R \gg R_0$, where in this case $R_0$ is the present day Ricci scalar. Further theoretical constraints must be imposed on $f(R)$ models to ensure that they exhibit a physically viable sequence of cosmological epochs. We call such models viable $f(R)$ model and details on those constraints are discussed in Sec.~\ref{sec: Perturbations}. However, any background that is compatible with a viable $f(R)$ model can be specified. Additionally, the existence of the scalaron results in more freedom for the background of $f(R)$ models. Thus, background cosmological evolution alone does not suffice in distinguishing between different $f(R)$ models including $\Lambda$CDM. Therefore, studying cosmological perturbations breaks the degeneracy.

In this work, we shall consider two of the most paradigmatic $f(R)$ mo\-dels widely considered in the literature namely, the Hu \& Sawicki \cite{Hu:2007nk} and Starobinsky \cite{Starobinsky:2007hu} models. We adopt a linear theory of perturbations which allows us to derive a second order differential equation that governs the evolution of matter density perturbations. Finding its solution allowed us to study various quantities that govern the evolution of large-scale structures. Those quantities include the growth of the structure function $f\sigma_8$ which is a product of the growth rate $f$ and the amplitude of matter fluctuations $\sigma_8$. The latter two variables are the observables of the large-scale structure (LSS) surveys. However, due to the degeneracy that is seen between the galaxy bias factor $b$ and $\sigma_8$, LSS surveys have not been in general capable of measuring $f$ and $\sigma_8$ separately. Nonetheless, when additional use of galaxy-galaxy lensing is at hand to obtain the constraints on $b$, the aforementioned degeneracy is broken \cite{Song:2008qt}.  

In the following the two $f(R)$ models mentioned above will be thus confronted with updated datasets of BAO \cite{DESI:2025zgx}, Planck18 CMB shift parameters \cite{Planck:2018vyg}, and SNeIa from the Pantheon+ compilation \cite{Pan-STARRS1:2017jku}, together with a compilation of $f\sigma_8$ measurements, and moreover, to three data points for $f$ and $\sigma_8$ which we also use to obtain constraints on the values of the models parameters. In particular, we use the Monte-Carlo Markov chain (MCMC) analysis to provide constraints on the parameters that measure the deviations of $f(R)$ models away from the $\Lambda$CDM model, namely $b_\mathrm{HS}$ for Hu \& Sawicki model and $b_\mathrm{S}$ for Starobinsky model, see Sec. \ref{sec: Background}. Further constraints are performed on the usual cosmological parameters such as, $\Omega_\mathrm{m,0}, \sigma_{8,0}$, and $h$. For comparison purposes, we also report constraints on the parameters of the Dark-Energy-Spectroscopic-Instrument (DESI) dark-energy equation of state, namely $\omega_0$ and $\omega_a$ \cite{DESI:2025zgx}. 

As mentioned we will study the linear theory of cosmological perturbations. Within this framework, modifications to gravity generally introduce both time and scale dependence into the effective gravitational interaction. This behavior can be conveniently characterized by the effective gravitational coupling, $\mu$, also referred to as the effective Newton's constant, $G_\mathrm{eff}/G_\mathrm{N}$, where $G_\mathrm{N}$ is the bare Newton's constant. The quantity $\mu$ enters directly into the equation governing the growth of matter density perturbations by modifying the standard Poisson equation. 

In addition, deviations from GR can be characterized by the gravitational slip parameter, $\gamma$, and the light-deflection parameter, $\Sigma$. The former quantifies any non-standard deviations between the gravitational potential $\Phi$ and curvature potential $\Psi$. The latter is particularly relevant for testing and constraining modified gravity models through galaxy-galaxy lensing, as it is directly related to the Weyl potential, $\Phi+\Psi$. As such, we construct two-dimensional phase spaces, $\mu-1$ versus $\gamma-1$ and $\mu-1$ versus $\Sigma-1$ to provide a framework for formulating null-hypothesis tests of the Hu \& Sawicki and Starobinsky $f(R)$ models, with $\Lambda$CDM adopted as the base model. With sufficiently precise measurements from future surveys, these tests, when performed at different redshifts, could therefore provide a means of ruling out the considered $f(R)$ models.

The paper is organized as follows, in Sec.~\ref{sec: Background} we derive the relevant equations for the expansion history of the  $f(R)$ models. In Sec.~\ref{sec: Perturbations} we present the equations go\-verning the evolution of the matter density perturbation, as well as the resulting expressions for $\mu, \gamma, \text{ and } \Sigma$ in the $f(R)$ context. In Sec.~\ref{sec: Constraints} we provide the constraints on the model parameters followed by the results of the best-fit parameters based on the MCMC constraints in Sec.~\ref{sec: MCMC results}. In Sec.~\ref{sec: LSS functions} we provide a model dependent null hypothesis test of the $f(R)$ models. Finally, Sec.~\ref{sec: conclusion} summarizes our main results and conclusions.

Unless otherwise stated, we use natural units (i.e., $\hbar = k_B = 8\pi G_\mathrm{N}=1)$.

\section{Background Evolution of the $f(R)$ model}\label{sec: Background}

The $f(R)$ modified theories are usually represented by a gravitational Lagrangian with a general function of the Ricci scalar $R$, thus the total action yields
\begin{equation}\label{Action}
    S = \int \text{d}^{4}x\sqrt{-g}\left[\frac{1}{2}f(R)+\mathcal{L}_\mathrm{M} \right] ,
\end{equation}
where $g$ is the determinant of the metric tensor $g_{\mu \nu}$, and $\mathcal{L}_\mathrm{M}$ is the Lagrangian of the matter content. It can be shown straightforwardly that by varying the action in Eq.~\eqref{Action} with respect to the metric $g_{\mu \nu}$ one can obtain the metric field equations as follows, 
\begin{equation}\label{Mod Eqns}
    R_{\mu \nu}f_{R}-\frac{1}{2}g_{\mu \nu}f(R) + \left(g_{\mu \nu}\Box - \nabla_{\mu}\nabla_{\nu}\right)\,f_{R} = T^{(\mathrm{\mathrm{M}})}_{\mu \nu}
\end{equation}
where $\Box \equiv \nabla_{\alpha}\nabla^{\alpha}$ and $f_{R}\equiv \text{d} f(R)/\text{d} R$, while, $T^{(\mathrm{M})}_{\mu \nu}$ is the energy-momentum tensor of the matter fields defined as,
\begin{equation}
    T^{(\mathrm{M})}_{\mu \nu} \equiv -\frac{2}{\sqrt{-g}}\frac{\delta \mathcal{L}_{\mathrm{M}}}{\delta g^{\mu \nu}},
\end{equation}
which satisfies the Bianchi identity $\nabla^{\mu}T_{\mu \nu}^{(\mathrm{M})} = 0$. 

We primarily consider a spatially flat Friedmann–Lemaître–Robertson–Walker (FLRW) line element in the comoving Cartesian coordinates written as,
\begin{equation}\label{FLRW metric}
    \text{d}s^{2}=-\text{d}t^{2}+a(t)^{2}(\text{d}x^{2}+\text{d}y^{2}+\text{d}z^{2}).
\end{equation}
Therefore, the corresponding modified Friedmann's equations in the matter dominated epoch can be derived from field equations Eq.~\eqref{Mod Eqns} once Eq.~(\ref{FLRW metric}) is considered. Thus,
\begin{align}\label{Cosmo Eqns}
    H^2 &= \frac{1}{3f_{R}}\left(\rho_\mathrm{m} +\frac{Rf_R-f}{2}-3H\dot{R}f_{RR}\right)\,, \notag \\
    -3H^2 - 2\dot{H} &= \frac{1}{f_R}\left[\dot{R}^2 f_{RRR} +(2H\dot{R}+\ddot{R})f_{RR}+\frac{1}{2}(f-Rf_R)\right],
\end{align}
where a subscript $R$ denotes derivatives with respect to the Ricci scalar $R$, the dot denotes a derivative with respect to cosmic time ($t$) with $H(t)\equiv \dot{a}/a$ being the Hubble rate, and $\rho_\mathrm{m}$ denotes the matter energy density. Note that $a$ is the scale factor and is defined in terms of redshift as $a \equiv 1/(1+z)$. In the background as given by Eq.~(\ref{FLRW metric}), the Ricci scalar satisfies the following expression,
\begin{equation}
    R \equiv 6\left(\dot{H}+2H^2\right).
    \label{Ricci}
\end{equation}

The continuity equation for matter (dust), when written as,
\begin{equation}\label{MatterD_equan}
    \dot{\rho}_\mathrm{m} + 3H\rho_\mathrm{m}=0,
\end{equation}
was utilized to reduce the number of independent equations. Note that the straightforward integration of Eq.~(\ref{MatterD_equan}) gives $\rho_\mathrm{m}=\Omega_\mathrm{m,0}\,a^{-3}$.

We will follow the formalism that was implemented in the following  \cite{de_la_Cruz_Dombriz_2016, Carloni_2007, Abdelwahab_2012} where the cosmological equations in Eq.~(\ref{Cosmo Eqns}) are expressed as a set of first-order autonomous system of equations in order to study the evolution of the cosmic background expansion history of a general class of $f(R)$ theories.
For instance, in \cite{de_la_Cruz_Dombriz_2016}, those equations are rewritten in terms of the following dimensionless variables,
\begin{align}\label{DynamicVariables}
    x \equiv \frac{\dot{R}f_{RR}}{f_RH}, \quad\quad\quad & y \equiv \frac{R}{6H^2},\quad & \chi \equiv \frac{f}{6f_RH^2}, \notag \\
    \Omega \equiv \frac{\rho_\mathrm{m}}{3f_RH^2}, \quad\quad & h \equiv \frac{H}{H_0}\,.
\end{align}
Then the Eqs.~(\ref{Cosmo Eqns}) and (\ref{MatterD_equan}) are substituted back into the natural logarithmic derivative of $a$ ($N=\text{ln}(a)$) of the dynamical variables Eq.~(\ref{DynamicVariables}) to produce the following system of equations,
\begin{align}\label{eq: System_equans}
    \frac{\text{d}x}{\text{d}N} &= -\left[x^2+x(y+1)-2y+4\chi-\Omega\right]\,, \notag \\
    \frac{\text{d}y}{\text{d}N} &= -y\left(2y-xQ-4\right)\,, \notag \\
    \frac{\text{d}\chi}{\text{d}N} &= -\left[\chi(x+2y-4)-xyQ\right]\,, \notag\\
    \frac{\text{d}\Omega}{\text{d}N} &= -\Omega(x +2y-1)\,, \notag \\
    \frac{\text{d}h}{\text{d}N}&=-h(2-y)\,,
\end{align}
where $Q \equiv f_R/Rf_{RR}$ is the auxiliary variable allowing us to specify our $f(R)$ model. It should be noted that the system of equations (\ref{eq: System_equans}) is completely closed with the Friedmann - constraint - equation expressed as 
\begin{equation}\label{Friedmann_eqn}
    y-\chi-x+\Omega=1.
\end{equation}
In order to solve the dynamical system Eqs.~(\ref{eq: System_equans}), we impose initial conditions at $z\sim10^3$, i.e., deep inside the matter dominated epoch. At such redshifts, the evolution of the Universe is well governed by the $\Lambda$CDM model.\footnote{Thus, at high-enough redshifts, the Friedmann equation would be given as
\begin{equation}\label{LCDM_eqn}
     H^2_{\Lambda \text{CDM}}(z) = H_0^2\left( \Omega_\mathrm{m,0}a^{-3}+1-\Omega_\mathrm{m,0}\right),
\end{equation}
where, $H_0$ and $\Omega_\mathrm{m,0}$ are the Hubble rate and the matter energy density parameter at present time ($t_0$), respectively.} Thus, we use this model to fix the initial conditions. For example, the initial condition for the dimensionless variable $\chi$ is given by,
\begin{align}
    \chi_\mathrm{in} = \frac{f(R_\mathrm{in})}{6f_R(R_\mathrm{in})H^2(z_\mathrm{in})}, 
\end{align}
where $R_\mathrm{in}$ is the value of the Ricci scalar at redshift $z_\mathrm{in}=10^3$ once Eq.~\eqref{LCDM_eqn} is substituted in Eq.~\eqref{Ricci}. Throughout this work, we will use $\Omega_\mathrm{m,0}=0.315$ as reported in \cite{Planck18} unless otherwise stated. 

The variable $Q$ in the dynamical system of Eqs.~(\ref{eq: System_equans}) will allow us to trace the background evolution history via $h(z)$ for a specific $f(R)$ model. We thus consider the so-called viable $f(R)$ gravity models. The viability of those models is ensured provided they satisfy the following conditions \cite{DeFelice:2010aj},
\begin{align}\label{eq: fR conditions_1}
    f_R > 0 \text{ and } f_{RR} > 0 \text{ for } R > R_0>0,
\end{align}
where $R_0$ is the present day Ricci scalar. Then from the perspective of observational cosmology, a viable $f(R)$ model must further satisfy the following \cite{Kumar:2023bqj},
\begin{align}
    f(R) \to R-2\Lambda, \text{ for } R \gg  R_0,
\end{align}
where $\Lambda$ is the cosmological constant. In Ref.~\cite{DeFelice:2010aj} it has been shown that the following condition must be satisfied for the $f(R)$ model to depict a late-time de Sitter solution,
\begin{align}
    0 < \left. \frac{Rf_{RR}}{f_R}\right|_{r} \text{ for } r=-\frac{Rf_R}{f} = -2.
\end{align}
We will write the viable $f(R)$ model as,
\begin{align}\label{eq: generic f(R)}
    f(R) = R - 2\Lambda\,y(R, b),
\end{align}
where $y(R, b)$ is called the deviation function. As will be seen below, With $f(R)$ models written as Eq.~\eqref{eq: generic f(R)}, it is easy to study the deviations away from GR. 

\subsection{Hu \& Sawicki $f(R)$ gravity}

The Hu \& Sawicki (HS) model is described by the Lagrangian \cite{Hu:2007nk},
\begin{equation}\label{eq: Hu&Sawicki}
    f_{\rm HS}(R) = R - m^2\frac{c_1(R/m^2)^n}{1+c_2(R/m^2)^n},
\end{equation}
where $c_1,c_2$ are the free parameters of the model, while $m \approx \Omega_\mathrm{m,0}H_0^2$. Again, $m$ and $n$ are positive constants with $n\in \mathbb{N}^+$.
Some algebraic manipulations entail that Eq.~(\ref{eq: Hu&Sawicki}) can be written as \cite{Arjona:2018jhh},
\begin{align}\label{eq: simplfied H&S}
    f_{\rm HS}(R) &= 
    R - \frac{2\Lambda_{\text{HS}}}{1 + \left(\frac{b_{\mathrm{HS}}\Lambda_{\text{HS}}}{R}\right)^n}.
\end{align}
with  $\Lambda_{\text{HS}} = m^2c_1/2c_2$ and $b_{\mathrm{HS}} = 2c^{(1-\frac{1}{n})}/c_1$.
It is easy to map Eq.~(\ref{eq: simplfied H&S}) into Eq.~(\ref{eq: generic f(R)}) and realize that $y(R, b_{\mathrm{HS}}) = \left[1 + \left(\frac{b_{\mathrm{HS}}\Lambda_{\text{HS}}}{R}\right)^n\right]^{-1}$. Moreover, we see that when $b_{\mathrm{HS}} \to 0$ the function $y(R, b_{\mathrm{HS}}) \to 1$ and we recover the $\Lambda$CDM scenario. 

\subsection{Starobinsky $f(R)$ gravity}
The Starobinsky (S) model can be described by the following Lagrangian \cite{Starobinsky:2007hu},
\begin{equation}\label{eq: Starobinsky}
    f_{\rm S}(R) = R + \lambda R_*\left[\left(1 + \frac{R^2}{R_*^2}\right)^{-n}-1\right],
\end{equation}
where $R_*$ and $\lambda$ are the positive free parameters of the model while $n \in \mathbb{N}^+$. We then map Eq.~(\ref{eq: Starobinsky}) to the form as in Eq.~(\ref{eq: generic f(R)}) by defining the free parameters as \cite{Kumar:2023bqj},
\begin{align}
    \Lambda_{\rm S} = \frac{\lambda R_*}{2},  \quad\quad b_\mathrm{S} = \frac{2}{\lambda}\,
\end{align}
then,
\begin{align}
    f_{\rm S}(R) = R - 2\Lambda_{\rm S}\left\{1-\frac{1}{\left[1+\left(\frac{R}{b_{\mathrm{S}}\Lambda_{\rm S}}\right)^2\right]^n}\right\}.
\end{align}
Here $y(R,b_{\rm S})=1-1/\left[1+\left(\frac{R}{b_{\rm S}\Lambda_{\rm S}}\right)^2\right]^n
$. It is easy to see that for $b_{\rm S}\to0$, $y(R, b_{\rm S})\to1$. 

\section{Cosmological matter perturbations in $f(R)$ gravity}\label{sec: Perturbations}

It is well-known that the cosmic expansion history of the $\Lambda$CDM model, as given by Eq.~(\ref{LCDM_eqn}), is a special case of an enormous range of theoretical models, including several models of dynamical dark energy (DE) or even cases like a Taylor expansion of the equation of state (EoS) parameter $w(z)$, for instance the $w_0w_a$CDM, i.e., $w(a) = w_0 + w_a(1-a)$, where $w_0$ and $w_a$ are the free parameters of the model \cite{Chevallier:2000qy, Linder:2002et}. When $w_0 = -1$ and $w_a = 0$, the model exactly  mimics the $\Lambda$CDM model. Again, paradigmatic $f(R)$ models are able to mimic the $\Lambda$CDM expansion history \cite{Multamaki:2005zs}. 

Therefore, in order to efficiently study the departures of those models from the standard $\Lambda$CDM, a popular technique has consisted of considering the evolution of the first-order matter density perturbations and assess its statistical significance of departures of $\Lambda$CDM predictions when subject to comparison with LSS data. It is important to note that, for all numerical solutions presented in this work, we set $\Lambda_{\rm HS}=\Lambda_{\rm S}=\Lambda$, where
$
\Lambda = 3H_0^2\left(1-\Omega_\mathrm{m,0}\right),
$
corresponding to the present-day cosmological constant. We also fix ($n=1$) for both the Hu \& Sawicki and Starobinsky models. This choice is motivated not only by its widespread adoption in the literature \cite{Kumar:2025mzo, Arjona:2018jhh}, but also by its consistency with theoretical expectations and its phenomenological viability \cite{Capozziello:2007eu}. In particular, the $n=1$ case has been shown to provide a good fit to the observational data, while higher values of ($n$) generally do not offer a significant improvement in the fit \cite{delaCruz-Dombriz:2015tye, Kandhai:2015pyr}.

Let us then consider the perturbed metric in the Newtonian (longitudinal) gauge as given by \cite{Tsujikawa:2007gd},
\begin{align}\label{eq: perturbed metric}
    \mathrm{d}s^2 = -(1+2\Phi)\mathrm{d}t^2
    +(1-2\Psi)a^2\delta_{ij}\mathrm{d}x^i\mathrm{d}x^j,
\end{align}
where, once in the Fourier space, $\Phi\equiv\Phi(a,k)$ and $\Psi\equiv\Psi(a,k)$ are the scalar metric potentials, and $k$ is the comoving wavenumber. For $f(R)$ gravity, the relation between the metric potentials is given by \cite{Tsujikawa:2007gd,delaCruzDombriz:2010xy}
\begin{equation}\label{eq: potential diff}
    \Phi-\Psi = -\frac{f_{RR}}{f_R}\delta R,
\end{equation}
where $\delta R = R-R_0$, with $R$ and $R_0$ denoting the Ricci scalars computed from the perturbed metric Eq.~(\ref{eq: perturbed metric}) and the background FLRW metric Eq.~(\ref{FLRW metric}), respectively. Eq.~(\ref{eq: potential diff}) shows that, unlike in GR, the two metric potentials generally differ, i.e., there exists an $f(R)$-induced anisotropic stress, in $f(R)$ gravity due to the additional scalar degree of freedom ($f_{RR}\neq0$). By contrast, in the standard $\Lambda$CDM model, where gravity is described by GR and other sources of  anisotropic stress are negligible, the metric potentials turn out to be identical.
%
%
Provided that one works in the sub-horizon approximation, i.e., $k \gg aH$, the evolution of the matter density contrast, defined as
\begin{equation}
    \delta_\mathrm{m} \equiv \frac{\Delta\rho}{\rho_0} \equiv \frac{\rho - \rho_0}{\rho_0},
\end{equation}
where $\rho_0$ is the background matter density, is given by,
\begin{align}\label{eq: density contrast}
    \ddot\delta_\mathrm{m}(a) + \left(\frac{\dot H(a)}{H(a)}+\frac{3}{a}\right)\dot\delta_\mathrm{m}(a)-\frac{3}{2}\frac{\Omega_\mathrm{m,0}\mu }{a^5H(a)^2/H_0^2}\delta_\mathrm{m}(a) = 0,
\end{align}
where a dot denotes the derivative with respect the scale factor
and 
$\mu\equiv G_{\rm eff}/G_\mathrm{N}$ holds for the dimensionless effective gravitational coupling governing the motion of gravi\-ta\-ting bodies. To solve Eq.~\eqref{eq: density contrast}, the background $H(a)$ is obtained by numerically solving Eq.~\eqref{eq: System_equans}. 

%
%
%

For $f(R)$ gravity, one can write the analytical expressions for $\mu$ and $Q_\mathrm{eff}$ as \cite{Tsujikawa:2007gd,Arjona:2018jhh}
\begin{align}\label{eq: Geff&Qeff}
    \mu =\frac{1}{f_R}
    \frac{1+4\dfrac{k^2}{a^2}\dfrac{f_{RR}}{f_R}}
         {1+3\dfrac{k^2}{a^2}\dfrac{f_{RR}}{f_R}}\,,
    \; 
    Q_{\rm eff}
    =\frac{1}{f_R}
    \frac{1+2\dfrac{k^2}{a^2}\dfrac{f_{RR}}{f_R}}
         {1+3\dfrac{k^2}{a^2}\dfrac{f_{RR}}{f_R}}\,.
\end{align}

\begin{figure*}[!t]
    \centering
    \includegraphics[width = 0.49\textwidth]{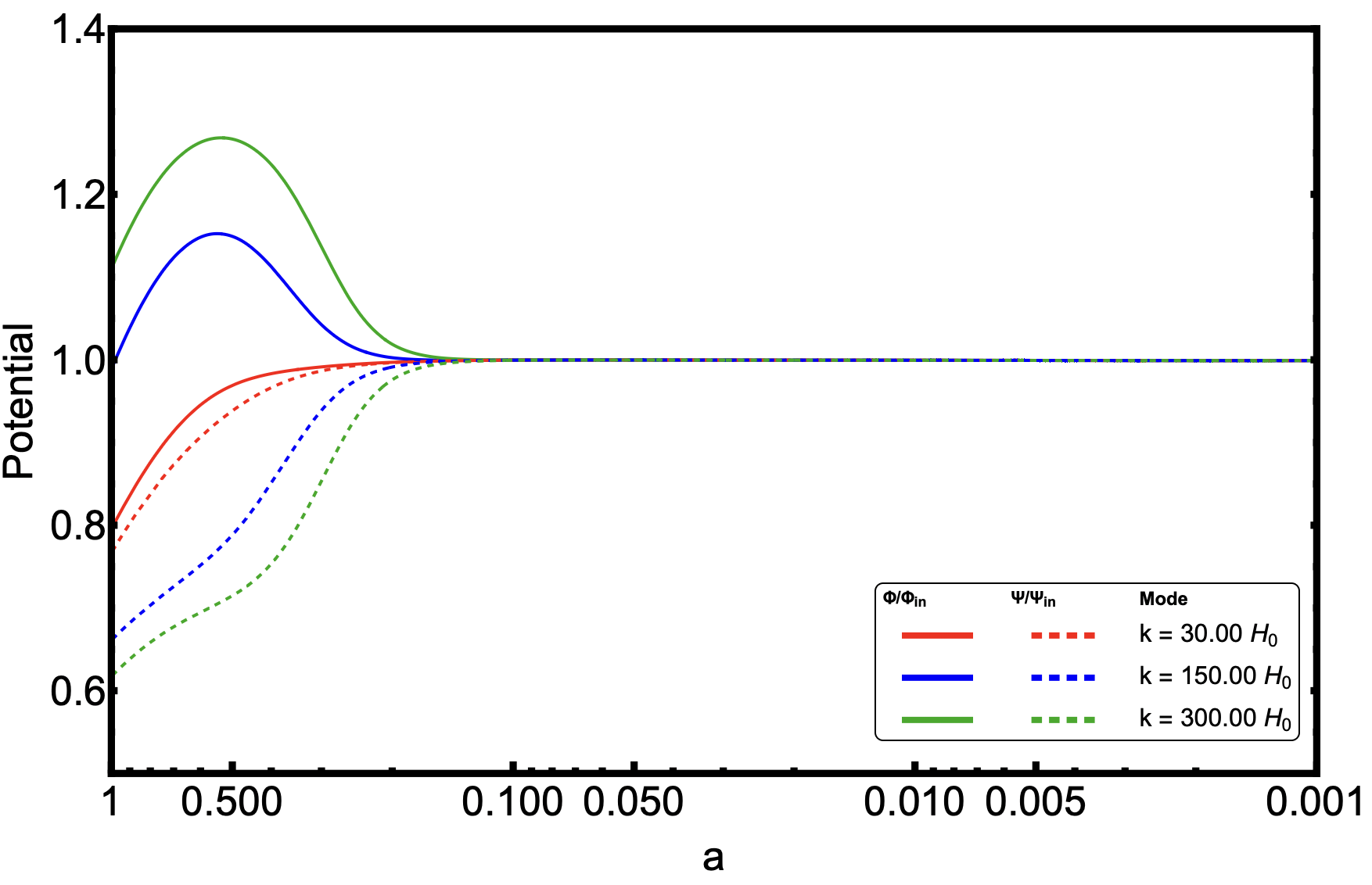}
    \hfill
    \includegraphics[width = 0.49\textwidth]{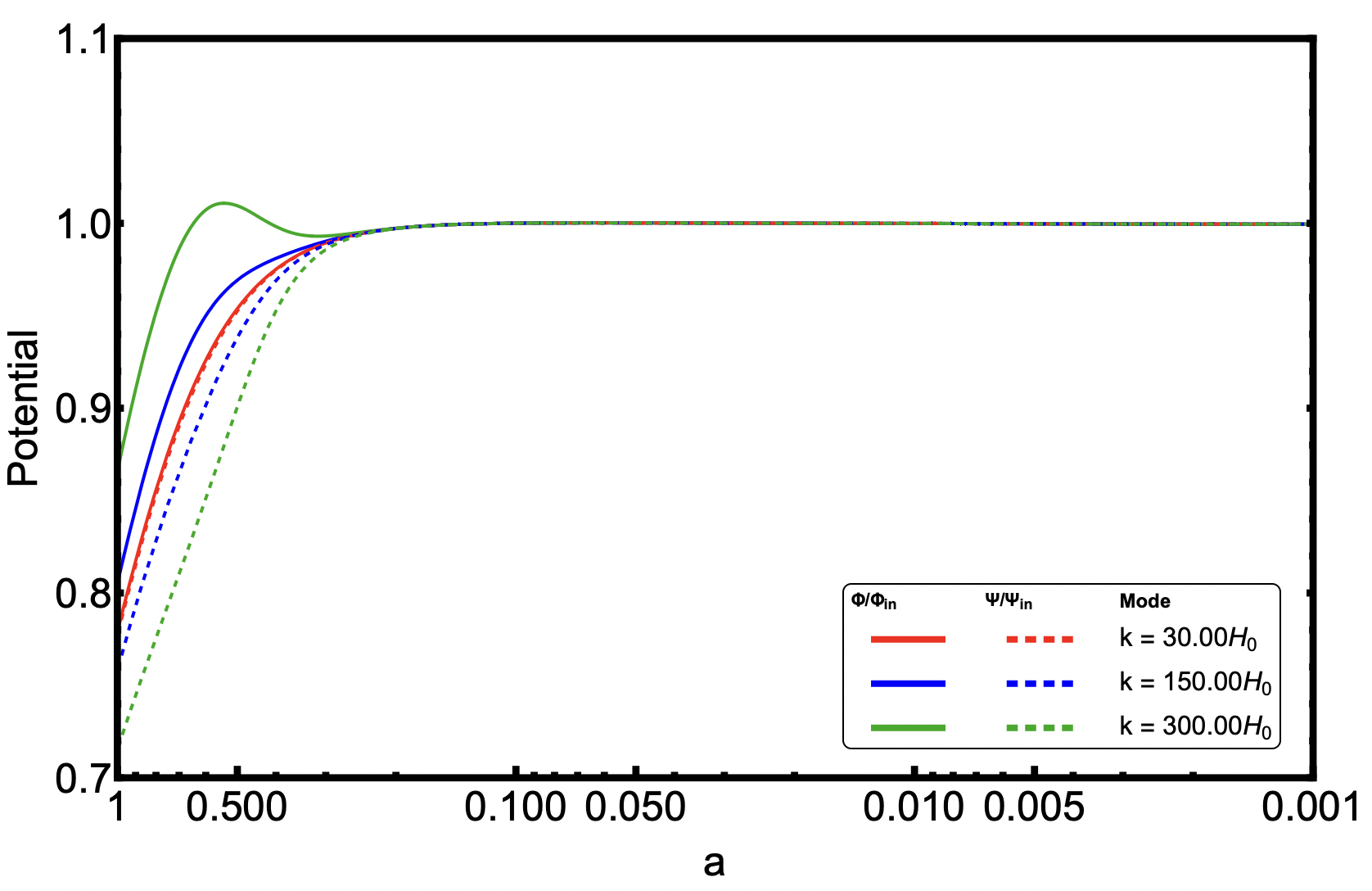}
    \caption{The evolution of the normalized potentials $\Phi/\Phi_{in}$ (solid) and $\Psi/\Psi_{in}$ (dashed) for the Hu \& Sawicki (left panel) and Starobinsky (right panel) models. For Hu \& Sawicki model we use $b_{\rm HS} = 10^{-3}$ and for Starobinsky model we use $b_{\rm S}=10^{-2}$. The green, blue, and red pairs of curves correspond to modes $k=30H_0, 150H_0,$  and $300H_0$ respectively. At late times, a clear divergence between the potentials emerges as a consequence of non-zero $f(R)$-induced anisotropic stress during these epochs. }
    \label{fig: Potentials}
\end{figure*}
It is important at this point to study the evolution of matter perturbations in the $f(R)$ gravity models. This can be done using the linear growth rate $f$ and the amplitude of matter fluctuations $\sigma_8$ which are related to the matter density contrast $\delta_\mathrm{m}$ by the following equations,
\begin{align}\label{eq: Growth rate}
    f(z) = -(1+z)\frac{\delta_\mathrm{m}'}{\delta_\mathrm{m}}, \quad \sigma_8(z) = \sigma_{8,0}\frac{\delta_\mathrm{m}}{\delta_0}
\end{align}
here the prime is the derivative with respect to redshift. Due to the existence of the degeneracy between the galaxy linear bias and the amplitude of matter fluctuation (i.e., the $b_1-\sigma_8$ degeneracy) the redshift-space distortion (RSD) surveys usually measure the following quantities \cite{Euclid:2025bxg, Song:2008qt},
\begin{align*}
    \quad b_1\sigma_8 \quad \text{and}\quad \beta = f/b_1
\end{align*}
Thus, without the best-fit constraint of $b_1$, only the quantity $f\sigma_8$ can be measured.

\begin{figure}
    \centering
    \hspace*{-1cm}
    \includegraphics[width=1.0\linewidth]{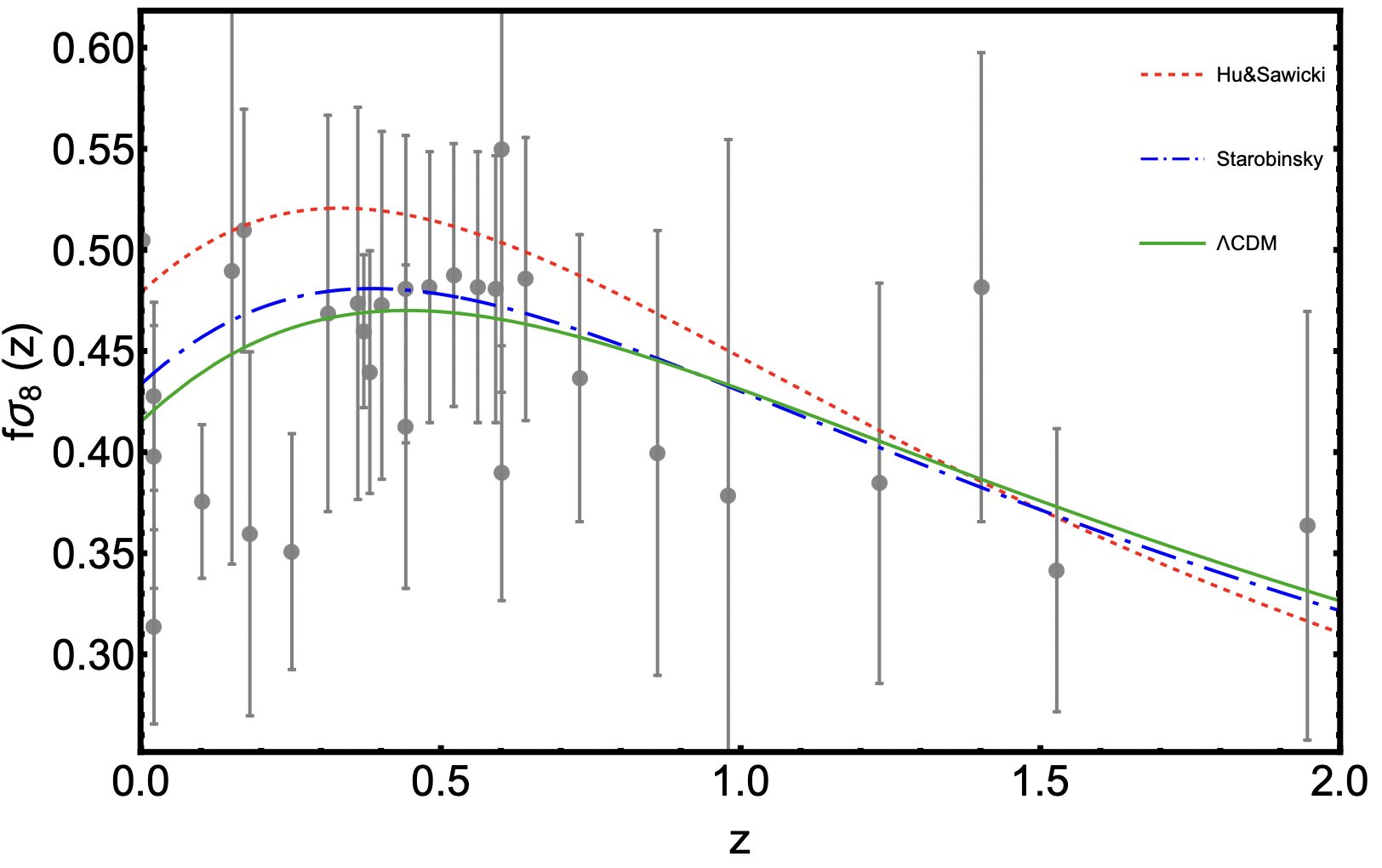}
    \caption{The redshift evolution of the growth of structure function $f\sigma_8$ for the Hu \& Sawicki model with $b_{\rm HS} = 10^{-3}$ and Starobinsky model with $b_{\rm S}=0.01$ for $k = 300H_0$. We also show the compilation of the redshift-space distortion (RSD) data from different surveys as shown in Table \ref{tab:fsigma_f_sigma_data}. The dotted grey curve corresponds to the standard $\Lambda$CDM model.}
    \label{fig:fsigma8}
\end{figure}

In Table \ref{tab:fsigma_f_sigma_data} we show a compilation of $f\sigma_8$ data obtained from various RSD surveys. In particular, we will restrict the data to only the robust and independent collection obtained in \cite{Nesseris:2017vor, Perenon:2019dpc}. In Fig.~\ref{fig:fsigma8} we reproduce the nu\-me\-ri\-cal solutions of the function $f\sigma_8$ for the Hu \& Sawicki, Starobinsky and the $\Lambda$CDM models and the data in Table \ref{tab:fsigma_f_sigma_data} is also shown.

\begin{figure*}[!t]
    \centering
    \includegraphics[width = 1\textwidth]{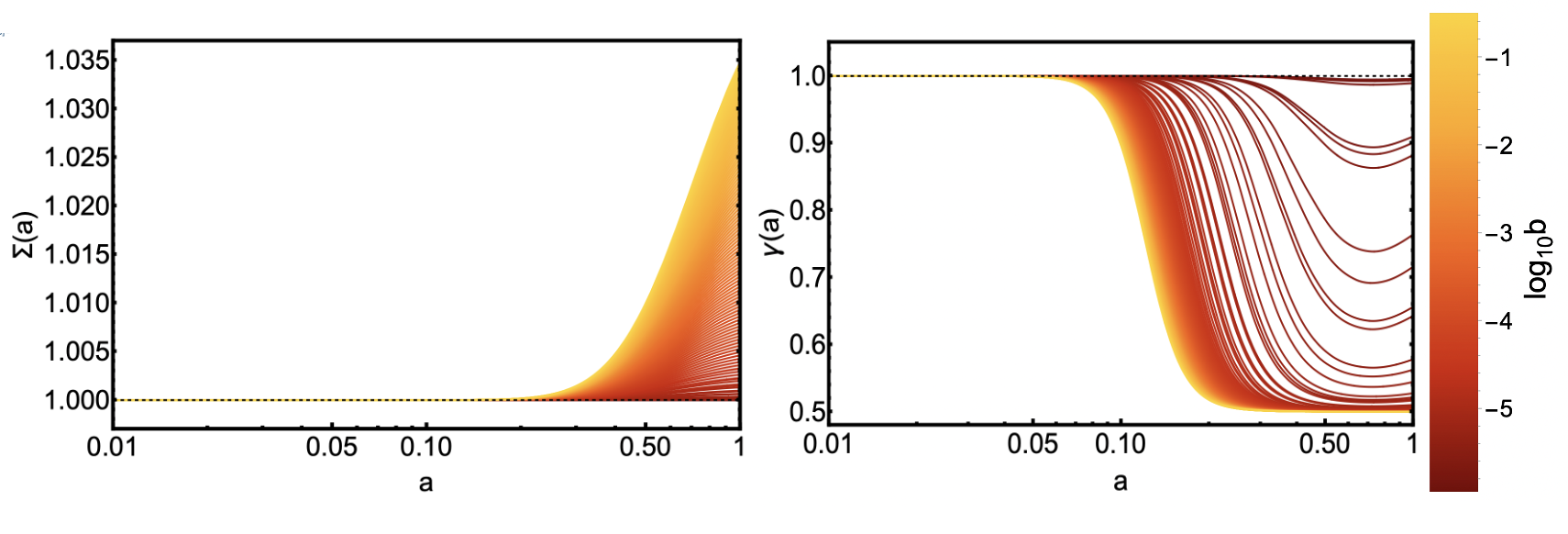}
    \caption{The evolution of the light deflection parameter $\Sigma$ and the slip parameter $\gamma$ for various values of $\text{log}_{10}b$ $ \in \{-6, -1\}$ for the Hu \& Sawicki model. The dotted black line correspond to the $\Lambda$CDM scenario (i.e $\Sigma = 1$, $\gamma=1$) and we considered the mode $k = 300H_0$ to reproduce the results. The deviations from the standard scenario are seen at $a \sim 0.183$ and $a \sim 0.050$ for $\Sigma$ and $\gamma$ respectively. Notice that $\gamma$ approaches $1/2$ at late-times while $\Sigma$ seem to not be bounded above.}
    \label{fig: Gamma&Sigma_H&S}
\end{figure*}

\begin{figure*}
    \centering
    \includegraphics[width = 1\textwidth]{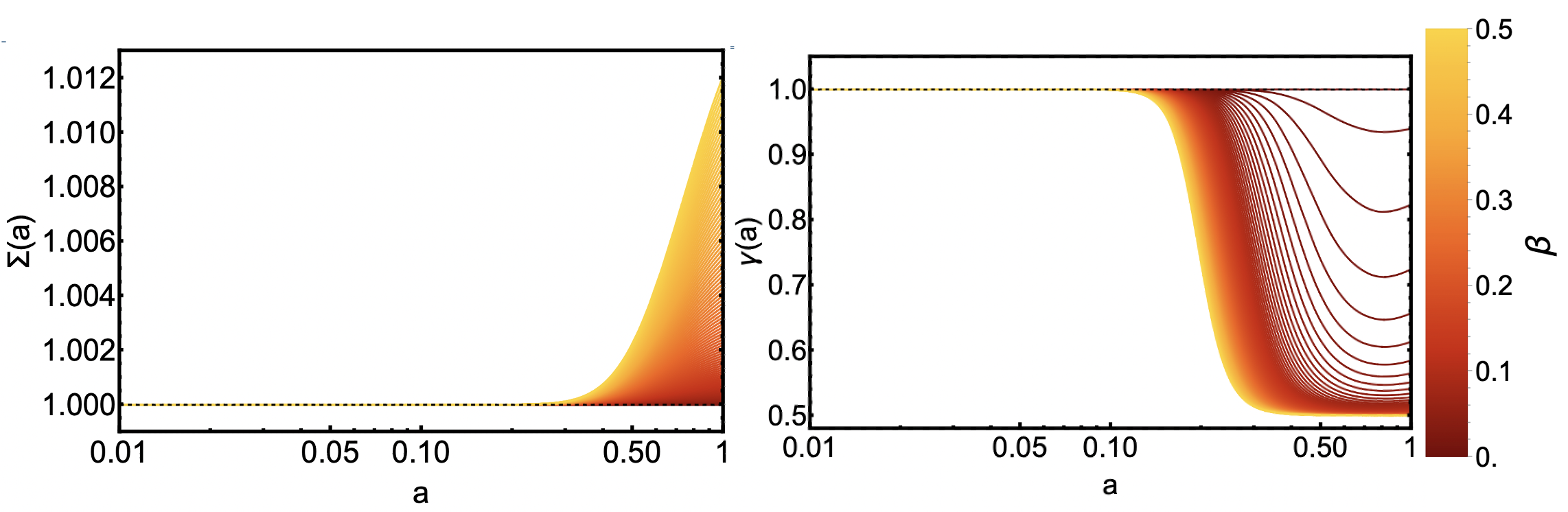}
    \caption{The evolution of the gravitational light deflection parameter $\Sigma$ and the slip parameter $\gamma$ for various values of $\beta$ $ \in \{0.001, 0.6\}$ for the Starobinsky model. The dotted black line correspond to the $\Lambda$CDM scenario (i.e $\Sigma = 1$, $\gamma=1$) and we considered the mode $k = 300H_0$ to reproduce the results. The deviations from the standard scenario are seen at $a \sim 0.105$ and $a \sim 0.246$ for $\Sigma$ and $\gamma$ respectively. Again at late-times it can be seen that $\gamma \to 1/2$ and $\Sigma$ is not bounded above.}
    \label{fig: Gamma&Sigma_Starobinsky}
\end{figure*}

Also, it can be shown that, under the sub-horizon approximation, the metric potentials satisfy the modified Poisson equations \cite{Murakami:2023qdl,Orjuela-Quintana:2023zjm,Tsujikawa:2007gd}
\begin{eqnarray}
\label{eq: potentions definition}
\frac{k^2}{a^2}\Phi &=& -4\pi G_\mathrm{N} \mu\rho_\mathrm{m}\delta_\mathrm{m},\\
\frac{k^2}{a^2}\Psi &=& -4\pi G_\mathrm{N} Q_{\rm eff}\rho_\mathrm{m}\delta_\mathrm{m}.
\end{eqnarray}
In Fig.~\ref{fig: Potentials} we show the evolution of the potentials $\Phi$ and $\Psi$ for both the Hu \& Sawicki and Starobinsky models for several sub-Hubble $k$ modes.
The potentials $\Phi$ and $\Psi$ can be used to derive other useful cosmological observables such as the gravitational slip parameter $\gamma \equiv \Phi/\Psi$. This parameter quantifies any non-standard relation between $\Phi$ and $\Psi$ and for $f(R)$ models it takes the form \cite{Tsujikawa:2007gd, Du:2026cly}\footnote{Note that in the latter reference, the convention for the parameter $\gamma$ is $\eta = \Phi/\Psi -1$.},
\begin{equation}\label{eq: gamma}
    \gamma = \frac{1+2\frac{k^2}{a^2}\frac{f_{RR}}{f_R}}{1+4\frac{k^2}{a^2}\frac{f_{RR}}{f_R}}\,.
\end{equation}
Thus, with the gravitational slip as defined above and the effective gravitational coupling as per Eq. \eqref{eq: potentions definition}, one can define the light deflection parameter $\Sigma$. As mentioned in the Introduction, $\Sigma$ is related to the sum of the gravitational and curvature potential (or Weyl potential) as,

\begin{equation}
    \frac{k^2}{a^2}\left(\Phi +\Psi\right) = -4\pi\Sigma G_{\rm N}\rho_{\rm m}\delta_{\rm m},
\end{equation}
and in terms of $\mu$ and $\gamma$ \cite{Du:2026cly},
\begin{equation}
    \Sigma = \mu\left(\frac{1+\gamma}{2}\right),
\end{equation}
where obviously $\Sigma_{\Lambda{\rm CDM}}\equiv1$.
In Figs.~\ref{fig: Gamma&Sigma_H&S} and \ref{fig: Gamma&Sigma_Starobinsky} we show the evolution of the LSS functions $\{\Sigma,\,\gamma\}$ for the Hu \& Sawicki model as per Eq.~(\ref{eq: Hu&Sawicki}), and for the Starobinsky model as per Eq.~(\ref{eq: Starobinsky}), respectively. The quantities $\{\Sigma, \gamma\}$ have similar evolutions for both models. However, the differences are seen when they deviate from the standard $\Lambda$CDM scenario.  

\section{Data and methodology}\label{sec: Constraints}
We perform an MCMC analysis to constrain the model parameters using a range of cosmological tracers for either the background or growth of large-scale structures. For the background expansion history, we use BAO, SNeIa , and CMB-derived background constraints. For the growth of structure, we use measurements of $f\sigma_{8}$, $f$, and $\sigma_8$. In particular, in this section we include an analysis of how the $\chi^2$ for each tracer mentioned above is derived. Then the MCMC analysis is performed using the following three combinations of $\chi^2$,
\begin{align}\label{eq: chiSq combinations}
\chi^2_{\mathrm{I}} &= \chi^2_{\mathrm{BAO}}+\chi^2_{\mathrm{SNeIa}}+\chi^2_{\mathrm{
CMB}}+\chi^2_{f\mathrm{\sigma_8}}, \notag \\
\chi^2_{\mathrm{II}} &= \chi^2_{\mathrm{BAO}}+\chi^2_{\mathrm{
SNeIa}}+\chi^2_{\mathrm{CMB}}+\chi^2_{
f}+\chi^2_{\mathrm{\sigma_8}}, \notag \\
\chi^2_{\mathrm{III}}&=\chi^2_{\mathrm{BAO}}+\chi^2_{\mathrm{SNeIa}}+\chi^2_{\mathrm{CMB}}+\chi^2_{f\mathrm{\sigma_8}}+\chi^2_{f}+\chi^2_{\mathrm{\sigma_8}}.
\end{align}
Note that the likelihoods $\mathcal{L}$ can be obtained from the $\chi^2$ as $\mathcal{L} \sim {\rm e}^{-\chi^2/2}$, assuming gaussianity and that the data are independent, thus the total likelihood is written as,
\begin{align}\label{eq: total L}
    \mathcal{L}_{\mathrm{tot}} = \mathcal{L}_{\mathrm{BAO}}\times\mathcal{L}_{\mathrm{SNeIa}}\times\mathcal{L}_{\mathrm{CMB}}\times\mathcal{L}_{\mathrm{growth}},
\end{align}
where, $\mathcal{L}_{\mathrm{growth}}$ is the likelihood constituting the growth of structure tracers \{$f\sigma_8, f, \sigma_8$\}.

\subsection{BAO Likelihood}
The BAO provide us with a powerful tool to measure the expansion history of the Universe, by employing a characteristic scale that appears in the matter clustering by pressure waves that propagate in a tightly coupled baryon-photon fluid before the last-scattering epoch. In this work we construct a BAO likelihood using a set of 13 data points coming from the DESI DRII survey \cite{DESI:2025zgx} that spans a redshift range $0.296 \leq z \leq 2.330$. The data vector is defined as,
\begin{align}
    \mathbf{V}_{\text{obs}} \equiv \{V_{j}^{\rm obs}\}_{j}^{13},
\end{align}
where each element of $\mathbf{V_{obs}}$ can correspond either to the isotropic BAO measurement,
\begin{align}
    V_{\text{iso}} \equiv \frac{D_{\rm V}(z)}{r_{\text{drag}}},
\end{align}
or to an anisotropic BAO measurement,
\begin{align}
    V_{\text{ani}} \equiv \frac{D_{\rm M}(z)}{D_{\rm H}(z)}.
\end{align}
Here, $D_{\rm V}(z)$ is the volume-averaged distance defined as,
\begin{align}
    D_{\rm V}(z) \equiv \left[(1+z)^2D_{\rm M}^2(z)\frac{cz}{H(z)}\right]^{1/3},
\end{align}
where $D_{\rm H}(z)\equiv c/H(z)$ is the Hubble distance and $H(z)$ is the Hubble rate and $D_{\rm M}(z)$ is the transverse comoving distance given by $D_{\rm M}\equiv(1+z)D_{\rm A}$ where $D_{\rm A}$ is the angular diameter distance,

\begin{equation}
    D_{\rm A} = \frac{1}{1+z}\int_{0}^{z}\frac{c{\rm d}z'}{H(z')}.
\end{equation}

While $r_{\rm drag}$ is the sound horizon at the drag redshift ($z_{\text{drag}}$) were the acoustic BAO waves are frozen-in because the photons are longer able to further \text{drag} baryons. The drag redshift can be written as \cite{Aizpuru:2021vhd},
\begin{align}
z_{\text{drag}} &= \frac{1}{0.71413\,(\Omega_{\rm m} h^2)} \times\notag \\
&\quad  
[1 + 428.169\, (\Omega_{\rm b} h^2)^{0.25646} (\Omega_{\rm m} h^2)^{0.61639} \\
  &+ 925.56\, (\Omega_{\rm m} h^2)^{0.75162}]
\end{align}

where the baryon density is fixed to a Planck18 value $\Omega_\mathrm{b,0}h^2 = 0.02218$ \cite{Planck:2018vyg} and the matter density $\Omega_\mathrm{m,0}h^2$ will be constrained.  
The sound horizon can be computed as,
\begin{align}
    r_{\text{drag}} = \int_{z_{\text{drag}}}^{\infty}\frac{c_{\rm s}(z)}{H(z)}{\rm d}z,
\end{align}
where $c_{\rm s}(z)$ is the sound speed of the baryon-photon fluid. 

Given a set of cosmological parameters $\theta$, we can write the BAO $\chi^2$ as,
\begin{align}\label{eq: BAO chi-sq}
    \chi^2_{\text{BAO}}(\theta) = \Delta \mathbf{V}^{\text{T}}(\theta)\mathbf{C}_{\text{DESI}}^{-1}\Delta \mathbf{V}(\theta),
\end{align}
where, $\Delta\mathbf{V}(\theta)=\mathbf{V}_{\text{theory}}-\mathbf{V}_{\text{obs}}$ is the residual vector and $\mathbf{V}_{\text{theory}}(\theta)$ is the theory vector. The covariance metric $\mathbf{C}_{\text{DESI}}$ is constructed as,
\begin{equation}
\mathbf{C}_{\rm DESI} =
\begin{pmatrix}
\sigma_{\rm BGS}^2 & 0 & \cdots & 0 \\
0 & \mathbf{C}_{\text{LRG}1} & \cdots & 0 \\
\vdots & \vdots & \ddots & \vdots \\
0 & 0 & \cdots & \mathbf{C}_{\text{Ly}\alpha}
\end{pmatrix},
\end{equation}
where $\sigma_{\text{BGS}}=0.075$ and the remaining $2\times2$ diagonal entries are the correlated pairs of the BAO tracer measurements. We adopt the following construction,
\begin{equation}
    \mathbf{C}^{(2)} = 
    \begin{pmatrix}
        \sigma_1^2 & \sigma_1\sigma_2\sigma_3 
        \\
        \sigma_1\sigma_2\sigma_3  & \sigma_2^2
    \end{pmatrix},
\end{equation}
where the $\sigma_j$ are the quoted errors for each of the BAO tracers.

\subsection{SNeIa Pantheon+ Likelihood}
We now provide the construction of the SNeIa likelihood using the Pantheon+ dataset with redshifts ranging between $0.01<z<2.261$ \cite{Pan-STARRS1:2017jku}. The observable is the corrected apparent magnitude, written in terms of the distance moduli,
\begin{align}
    \mu_{\text{theory}}(z) = 5\text{log}_{10}\left[\frac{1+z_{\text{hel}}}{1+z_{\text{CMB}}}\frac{D_\mathrm{L}(z)}{\text{Mpc}}\right] + 25,
\end{align}
where we have accounted for the difference between the heliocentric and CMB-frame redshifts. The luminosity distance for a flat universe ($\Omega_k = 0$) is given by,
\begin{align}
    D_{\rm L}(z)\equiv (1+z)\int_{0}^{z}\frac{c\,{\rm d}z'}{H(z')},
\end{align}
and the residual vector is defined as,
\begin{equation}
    \Delta m(z) = m_{\rm B}^{\rm corr} - \mu_{\rm theory}(z),
\end{equation}
where $m_{\rm B}^{\rm corr}$ are the corrected apparent magnitudes. The marginalized $\chi^2$ over the absolute magnitude $M_{\text{B}}$ of the Pantheon+ SNeIa Ia is,
\begin{equation}\label{eq: SNeIa chi-sq}
    \chi_{\text{SNe}}^2(\theta) = \Delta m^{\text{T}}\textbf{C}^{-1}\Delta m - \frac{\left(\textbf{1}^{\text{T}}\textbf{C}^{-1}\Delta m\right)^2}{\textbf{1}^{\text{T}}\textbf{C}^{-1}\textbf{1}} + \text{ln}\left(\frac{\textbf{1}^{\text{T}}\textbf{C}^{-1}\textbf{1}}{2\pi}\right),
\end{equation}
where the $\textbf{1}$ is the unit matrix and $\textbf{C}^{-1}$ is the inverse covariance matrix which includes both the the statistical and systematic errors. 

\subsection{CMB Likelihood}
Using the Planck distance priors that compress the full CMB into a small set of parameters that are sensitive to the background, we will incoparate the constrains from the cosmic microwave background (CMB). The data vector can be written as,
\begin{align}
\textbf{V}_{\text{CMB}} = \begin{pmatrix}
        R \\
        \ell_\mathrm{a} \\
        \Omega_\mathrm{b}h^2
    \end{pmatrix},
\end{align}
where $R$ is the shift parameter and $\ell_\mathrm{a}$ is the acoustic scale which are respectively defined at the redshift at decoupling $z_{\mathrm{dec}}$ as,
\begin{align}
    R \equiv \sqrt{\Omega_\mathrm{m}H_0^2}\frac{D_\mathrm{M}(z_{\mathrm{dec}})}{c}\,,\; 
    \ell_\mathrm{a} \equiv \pi\frac{D_\mathrm{M}(z_{\mathrm{dec}})}{r_s(z_{\mathrm{dec}})}.
\end{align}
The redshift at decoupling $z_{\mathrm{dec}}$ is the redshift at which the photons last scattered from the baryons before freely transversing the universe and are therefore imprinted in the CMB. This redshift is then obtained from a fitted formula as \cite{Aizpuru:2021vhd},
\begin{align}
    z_{\mathrm{dec}} &= \frac{391.672(\Omega_\mathrm{m}h^2)^{-0.372296}+937.422(\Omega_\mathrm{b}h^2)^{-0.97966}}{(\Omega_\mathrm{m}h^2)^{-0.0192951}(\Omega_\mathrm{b} h^2)^{-0.93681}} \notag \\
    &+ (\Omega_\mathrm{m}h^2)^{-0.731631}.
\end{align}
We then adopt the Plank 2018 distance priors measurements  with their corresponding errors \cite{Planck:2018vyg},
\begin{equation}
    \textbf{V}_{\rm obs} = \begin{pmatrix}
        1.745 \pm 0.093 \\
        301.77 \pm 0.0051 \\
        0.02248 \pm 0.00016
    \end{pmatrix}.
\end{equation}
Therefore, the CMB $\chi^2$ is defined as,
\begin{equation}\label{eq: CMB chi-sq}
    \chi^2_{\rm CMB}(\theta) \equiv \Delta\mathbf{V}^{\rm T}\,\mathbf{C}^{-1}_{\rm CMB}\,\Delta\mathbf{V},
\end{equation}
where, as usual, $\Delta V\equiv \mathbf{V}_{\rm theory} - \mathbf{V}_{\rm obs}$ is the residual vector and $\mathbf{C}_{\rm CMB}^{-1}$ is the CMB inverse covariance matrix given by,
\begin{equation}
    \mathbf{C}_{\rm CMB} = 1\times10^{-8}\begin{pmatrix}
        2556.7782 & 23212.22 & -57.34582 \\
        23212.22 & 830122.02 & -628.5626 \\
        -57.34582 & -628.5626 & 2.530009
    \end{pmatrix}.
\end{equation}

\subsection{Growth Likelihood}
We then construct the $\chi^2$ for the quantities that probe the LSS which include $f\sigma_8$, $f$, and $\sigma_8$ respectively as follows,

\begin{align}
    \chi^2(\theta)_{f\sigma_8} = \Delta\textbf{X}^{\rm T}\textbf{C}^{-1}_{f\sigma_8}\Delta\textbf{X}, \\
    \chi^2(\theta)_{f} = \Delta\textbf{Y}^{\rm T}\textbf{C}_{f}^{-1}\Delta\textbf{Y}, \\
    \chi^2(\theta)_{\sigma_8} = \Delta\textbf{Q}^{\rm T}\textbf{C}_{\sigma_8}^{-1}\Delta\textbf{Q},
\end{align}
where the covariance matrices $\textbf{C}_{j}$ are obtained by diagonalizing the  squares of the errors as $\textbf{C}_j= \textit{\rm diag}(\sigma^2_1 \quad \sigma^2_2 \quad ... \quad \sigma^2_k)$ where $j = \{f\sigma_8, f, \sigma_8\}$ and $k =\{27, 3\}$. For a given set of cosmological parameters $\theta$, the residual vectors for the quantities $\in \{j\}$ are given by,
\begin{align}
    \Delta \textbf{X} = f\sigma_8(z_i, \theta) - (f\sigma_8)_{\rm obs}, \\
    \Delta \textbf{Y} = f(z_i, \theta) - (f)_{\rm obs},  \\
    \Delta \textbf{Q} = \sigma_{8}(z_i, \theta) - (\sigma_8)_{\rm obs},
\end{align}
where $z_i$ are the $i^{\rm th}$ observed redshifts and $(f\sigma_8)_{\rm obs}, (f)_{\rm obs}, (\sigma_8)_{\rm obs}$ are the observed data for the quantities in $\{j\}$ as shown in Table \ref{tab:fsigma_f_sigma_data}.

\section{MCMC results}\label{sec: MCMC results}
\begin{figure}
\centering
\includegraphics[width = 0.5\textwidth]{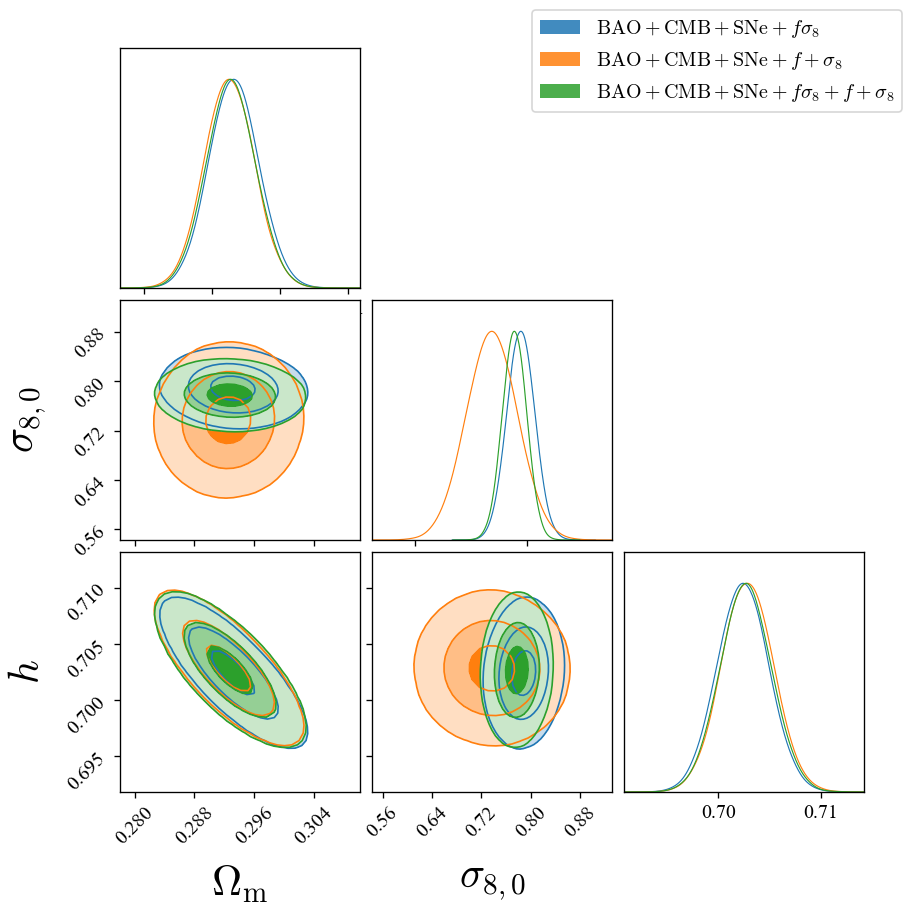}
\caption{The constraints for the parameters $\{\Omega_\mathrm{m,0}, h, \sigma_8\}$ for the  standard $\Lambda$CDM model. The three contour levels correspond to the $1\sigma, 2\sigma, \text{ and } 3\sigma$ interval levels.}
\label{fig:LCDM MCMC}
\end{figure}

\begin{figure}
\centering
\includegraphics[width = 0.5\textwidth]{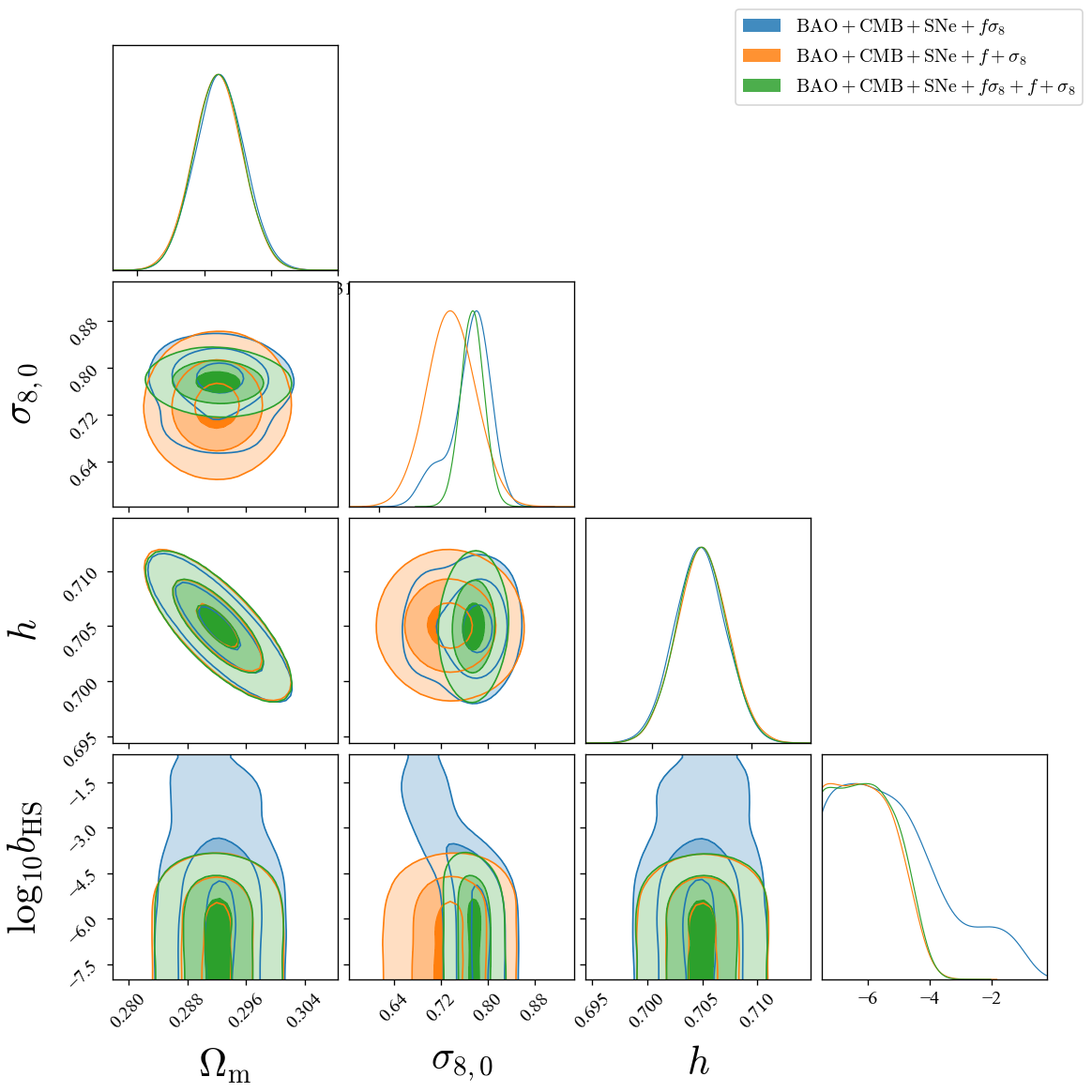}
\caption{Constraints for the parameters $\{\Omega_\mathrm{m,0}, h, \sigma_8, \text{log}_{10}b_{\rm HS}\}$ for the  standard Hu \& Sawicki model. The three contour levels correspond to the $1\sigma, 2\sigma, \text{ and } 3\sigma$ interval levels. }
\label{fig:H&S_MCMC}
\end{figure}

\begin{figure}
\centering
\includegraphics[width = 0.5\textwidth]{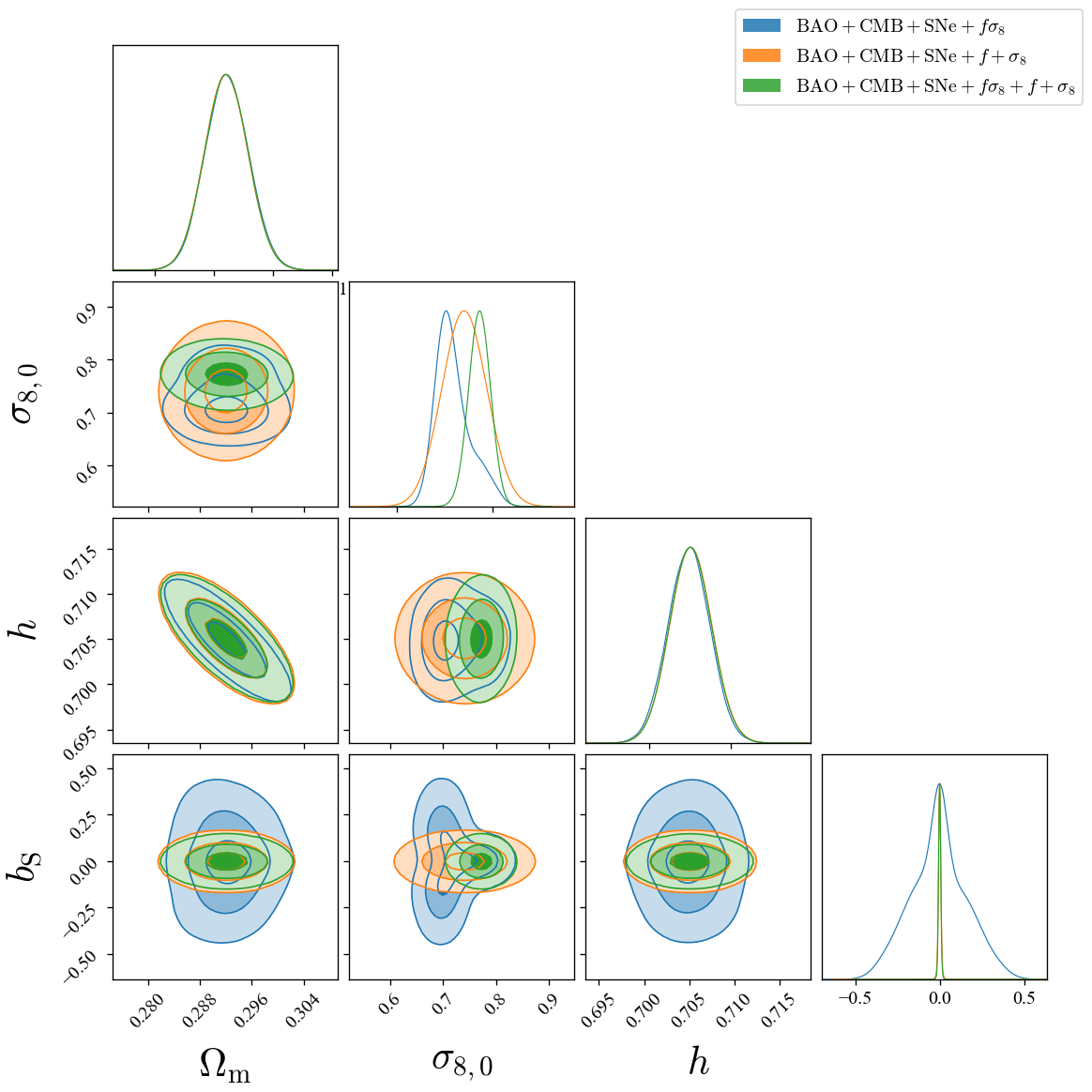}
\caption{The constraints for the parameters $\{\Omega_\mathrm{m,0}, h, \sigma_8, b_{\rm S}\}$ for the Starobinsky model. The three contour levels correspond to the $1\sigma, 2\sigma, \text{ and } 3\sigma$ interval levels.}
\label{fig:Starobinsky_MCMC}
\end{figure}

\begin{figure}
\centering
\includegraphics[width = 0.5\textwidth]{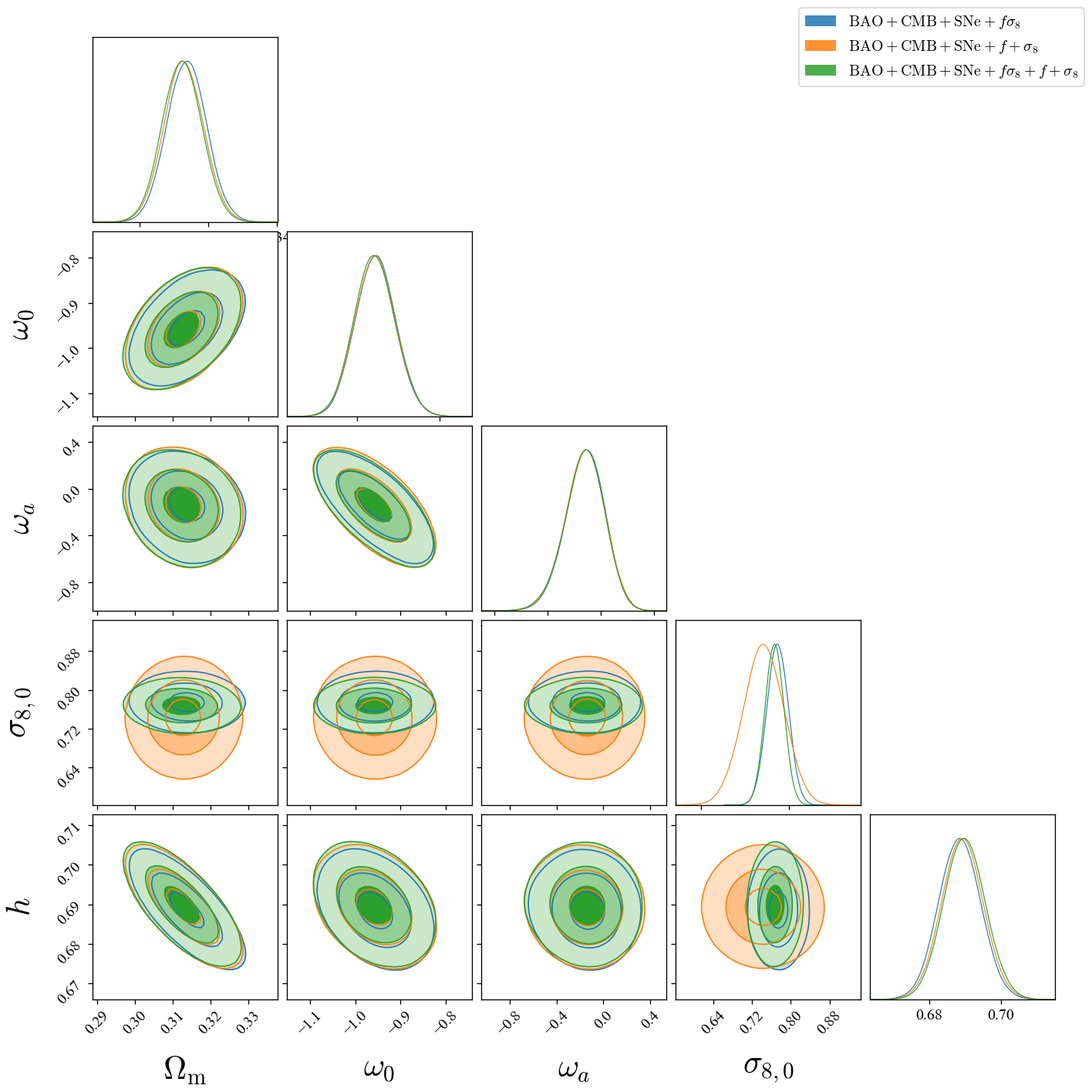}
\caption{The constraints for the parameters $\{\Omega_\mathrm{m,0}, h, \sigma_8\}$ for the $w_0w_a$CDM model. The three contour levels correspond to the $1\sigma, 2\sigma, \text{ and } 3\sigma$ interval levels.}
\label{fig:w0waCDM_MCMC}
\end{figure}

    In this section we provide the cosmological fits for Hu \& Sawicki, and the Starobinsky models by including the 1$\sigma$, 2$\sigma$ and 3$\sigma$ distributions for the combined datasets, BAO + CMB + SNeIa + $f\sigma_8$, BAO + CMB + SNeIa + $f + \sigma_8$, and BAO + CMB + SNeIa + $f\sigma_8 +f+\sigma_8$ as shown in Eq. (\ref{eq: chiSq combinations}). For comparison with benchmark models, we also displayed the corresponding results for both $\Lambda$CDM and $w_0w_a$CDM.  In particular, the contour levels for the $\Lambda$CDM model are depicted in Fig. \ref{fig:LCDM MCMC}, whereas the constraints for the Hu \& Sawicki and Starobinsky $f(R)$ models are shown in Fig. \ref{fig:H&S_MCMC} and Fig. \ref{fig:Starobinsky_MCMC}. Finally, the $w_0w_a$CDM model results are displayed in Fig. \ref{fig:w0waCDM_MCMC}.

To initiate the MCMC simulator we set the values for the parameters \{$\Omega_\mathrm{m,0}, h, \sigma_{8,0}$\} = \{$0.315\pm0.007, 0.674\pm0.005, 0.811\pm0.006$\}. Those values were taken from Ref.~\cite{Planck:2018vyg} and are reported with their $1\sigma$ errors respectively. While for the $w_0w_a$CDM model, the values for the parameters \{$w_0,w_a$\} are set to \{$-1, 0$\}. Lastly, the initial values for $\{\text{log}_{10}b_{\rm HS}, b_{\rm S}\}=\{-3, 0.01\}$ for the Hu \& Sawicki and Starobinsky models respectively.

For all combinations of datasets, we adopt the uniform wide flat priors on the parameters as presented in Table \ref{tab:priors}.

\begin{table}[H]
    \centering
    \begin{tabular}{c  c}
        \hline
        Parameter & Prior  \\
        \hline\hline
        $\Omega_\mathrm{m,0}$ & $\mathcal{U}[0.1, 0.6]$ \\
        
        $h$ & $\mathcal{U}[0.4, 1]$ \\

        $\sigma_{8,0}$ & $\mathcal{U}[0.2, 0.9]$ \\

        $w_0$ & $\mathcal{U}[-2, 1]$ \\

        $w_a$ & $\mathcal{U}[-3, 2]$ \\

        $\text{log}_{10}b_{\rm HS}$ & $\mathcal{U}[-8, 0]$ \\
        $b_{\rm S}$ & $\mathcal{U}[-0.5, 0.5]$ \\
        \hline
    \end{tabular}
    \caption{Priors imposed on the MCMC analysis in order to obtain the constraints on the parameters. The priors for $\Omega_\mathrm{m,0}, h$, and $\sigma_{8,0}$ where kept the same throughout all considered models, $\Lambda$CDM, $w_0w_a$CDM, Hu \& Sawicki and Starobinsky models.}
    \label{tab:priors}
\end{table}

\begin{table*}
\centering

\setlength{\tabcolsep}{6pt}
\renewcommand{\arraystretch}{1.3}

\resizebox{\textwidth}{!}{
\begin{tabular}{l c c c c c c c}
\hline
Parameters & $\Omega_\mathrm{m,0}$ & $\sigma_{8,0}$ & $h$ & $\text{log}_{10}b_{\rm HS}$ & $w_0$ & $w_a$ & $b_{\rm S}$ \\
\hline \hline

$\mathbf{\Lambda}$\textbf{CDM} &&&&&&& \\
BAO + CMB + SNeIa + $f\sigma_8$ 
& $0.2926\pm0.0036$ & $0.7881^{+0.0242}_{-0.0239}$ & $0.7047\pm0.0024$ & $-$ & $-$ & $-$ & $-$ \\

BAO + CMB + SNeIa + $f+\sigma_8$ 
& $0.2921\pm0.0036$ & $0.7338\pm0.0442$ & $0.7050\pm0.0024$ & $-$ & $-$ & $-$ & $-$ \\

BAO + CMB + SNeIa + $f\sigma_8 + f + \sigma_8$
& $0.2932\pm0.0035$ & $0.7763\pm0.0021$ & $0.7049\pm0.0024$ & $-$ & $-$ & $-$ & $-$ \\
\hline

\textbf{$w_0 w_a$}\textbf{CDM} &&&&&& \\
BAO + CMB + SNeIa + $f\sigma_8$
& $0.3137\pm0.0055$ & $0.7748\pm0.0287$ & $0.6885\pm0.0056$ & $-$ & $-0.9557\pm0.0473$ & $-0.1409^{+0.1707}_{-0.1808}$ & $-$ \\

BAO + CMB + SNeIa + $f+\sigma_8$
& $0.3128\pm0.0056$ & $0.7428^{+0.0451}_{-0.0443}$ & $0.6893\pm0.0056$ & $-$ & $-0.9569\pm0.0476$ & $-0.1440^{+0.0451}_{-0.0443}$ & $-$ \\

BAO + CMB + SNeIa + $f\sigma_8 + f+\sigma_8$
& $0.3123\pm{+0.0046}$ & $0.7678\pm0.0207$ & $0.6899\pm0.0056$ & $-$ & $-0.9590^{+0.0483}_{-0.0470}$ & $-0.1454^{+0.1706}_{-0.1867}$ & $-$ \\
\hline

\textbf{Hu \& Sawicki}  &&&&&&& \\
BAO + CMB + SNeIa + $f\sigma_8$
& $0.2926\pm0.0036$ & $0.7781^{+0.0286}_{-0.0449}$ & $0.7047\pm0.0025$ & $-5.6390^{+1.9602}_{-1.6157}$ & $-$ & $-$ & $-$  \\

BAO + CMB + SNeIa + $f+\sigma_8$
& $0.2921\pm0.0035$ & $0.7384^{+0.0443}_{-0.0446}$ & $0.7050\pm0.0025$ & $-6.3824^{+1.2035}_{-1.1071}$ & $-$ & $-$ & $-$  \\

BAO + CMB + SNeIa + $f\sigma_8 + f+\sigma_8$
& $0.2922\pm0.0035$ & $0.7764^{+0.2115}_{-0.0216}$ & $0.7050\pm0.0025$ & $-6.3250^{+1.2164}_{-1.1376}$ & $-$ & $-$ & $-$ \\
\hline

\textbf{Starobinsky} &&&&&&& \\
BAO + CMB + SNeIa + $f\sigma_8$
& $0.2921\pm0.0037$ & $0.7112^{+0.0439}_{-0.0268}$ & $0.7049\pm0.0025$ & $-$ & $-$ & $-$ & $0.0022^{+0.1650}_{-0.1703}$ \\

BAO + CMB + SNeIa + $f$ + $\sigma_8$ 
& $0.2920\pm0.0035$ & $0.7409\pm0.0452$ & $0.7051\pm0.0025$ & $-$ & $-$ & $-$ & $(0.7\pm62.0)\times10^{-4}$ \\

BAO + CMB + SNeIa + $f\sigma_8$ + $f$ + $\sigma_8$ 
& $0.2921\pm0.0036$ & $0.7723\pm0.0219$ & $0.7051\pm0.0025$ & $-$ & $-$ & $-$ & $(0.8\pm61.0)\times10^{-4}$ \\
\hline

\end{tabular}
}

\caption{Best-fit cosmological parameters $\Omega_\mathrm{m,0}$, $\sigma_{8,0}$, $h$, $\mathrm{log}_{10}b_{\rm HS}$, $b_{\rm S}$ $w_0$, and $w_a$ for the $\Lambda$CDM, $w_0w_a$CDM, Hu \& Sawicki and Starobinsky models.}
\label{tab1: bestfitparameters}

\end{table*}

\subsection{Model comparison}\label{sec: Comparison}
In this section, we proceed to compare statistically the $f(R)$ gravity models against the $\Lambda$CDM paradigm, with the latter taken as the benchmark model. A comparison between the $w_0w_a$CDM and $\Lambda$CDM models is also presented. The model comparison is performed using both the Akaike Information Criterion (AIC) \cite{Akaike} and the Bayesian Information Criterion (BIC), the latter also known as the Schwarz Information Criterion \cite{Schwarz}. These statistical criteria provide a means of assessing the relative performance of the $f(R)$ gravity models and $w_0w_a$CDM model with respect to the $\Lambda$CDM model. 

The AIC and BIC values for all the aforementioned models are obtained from the following expressions,
\begin{equation}
    \text{AIC} = \chi^2 + 2\text{K}\,,\;\;\;
    \text{BIC} = \chi^2 + \text{K}\,\text{log}(N_j) ,
\end{equation}
where $\chi^2$ is calculated from the minimum value of the Gaussian likelihood function $\mathcal{L}(\hat{\theta}|\mathcal{D}, \mathcal{M}) = \text{exp}[-\chi^2/2]$ of the model $\mathcal{M}$ given the dataset $\mathcal{D}$. While K is the number of the free parameters of the model and $N_j$ is the number of the data points used for the $j^{th}$ dataset.

The AIC and BIC are generally interpreted such that a smaller value indicates a preferred model. However, when comparing two or more models, the relative diffe\-ren\-ces between the information criteria provide a more meaningful measure of model preference. Additionally, the BIC penalizes models with extra degrees of freedom especially when the number of data points is sufficiently large. In this work, given that the models have different number of parameters, it will be crucial to provide comparison using both criteria. More specifically, we have 3 parameters for the $\Lambda$CDM and 5 parameters for $w_0w_a$CDM models. We have 4 parameters for both the Hu \& Sawicki and Starobinsky models.  

Hence, by taking the $\Lambda$CDM to be the reference model, the relative differences are therefore defined as $\Delta \text{AIC} = \text{AIC}_{\text{model}} - \text{AIC}_{\Lambda\text{CDM}}$, and $\Delta \text{BIC} = \text{BIC}_{\text{model}} - \text{BIC}_{\Lambda\text{CDM}}$. To interpret these differences, hereafter collectively denoted by $\Delta\mathrm{IC}_{(\mathrm{A,B})}$, we employ the Jeffreys' scale \cite{Nesseris:2012cq}. Thus, when $\Delta\mathrm{IC}{(\mathrm{A,B})}<0$, the model under consideration is preferred by the data over $\Lambda$CDM, whereas $\Delta\mathrm{IC}{(\mathrm{A,B})}>0$ indicates a preference for $\Lambda$CDM. The strength of the preference for one model over the other is determined by the magnitude of the relative difference. In particular,  $|\Delta\text{IC}_{(\text{A},\text{B})}| \le 2$ implies that the models being compared are statistically indistinguishable. While for $4\le|\Delta\text{IC}_{(\text{A},\text{B})}| \le 7$ implies a positive evidence and
$|\Delta\text{IC}_{(\text{A},\text{B})}| \geq 10$ implies an even stronger support. 

In Table \ref{tab2: comparison} we show the resulting values for $\mathcal{L}(\hat\theta|\mathcal{D}, \mathcal{M})$, $\chi^2$, $\chi^2-$reduced, AIC, $\Delta$AIC, BIC, and $\Delta$BIC for the models mentioned above. Given that for the combined datasets, $\Delta\text{AIC}<0$, we conclude that the $w_0w_a$CDM, Hu \& Sawicki and Starobinsky models are preferred over the $\Lambda$CDM model. However, the preference is less compelling when we consider the BIC. The latter fact is drawn from the fact that the magnitude of $\Delta$BIC is less than 2.

\begin{table*}
\centering

\setlength{\tabcolsep}{6pt}
\renewcommand{\arraystretch}{1.3}

\resizebox{\textwidth}{!}{
\begin{tabular}{l c c c c c c c}
\hline
Data & log$\mathcal{L}(\hat\theta|D)$ & $\chi^2$ & $\chi^2-$reduced & AIC & $\Delta \text{AIC}$ & BIC & $\Delta \text{BIC}$ \\
\hline \hline

$\mathbf{\Lambda}$\textbf{CDM} &&&&&& \\
BAO + CMB + SNeIa + $f\sigma_8$ 
& $-731.34$ & $1462.68$ & $0.897$ & $1468.68$ & $-$ & $1484.87$ & $-$ \\

BAO + CMB + SNeIa + $f+\sigma_8$ 
& $-728.23$ & $1456.46$ & $0.907$ & $1462.46$ & $-$ & $1478.61$ & $-$ \\

BAO + CMB + SNeIa + $f\sigma_8 + f + \sigma_8$
& $-735.83$ & $1471.66$ & $0.916$ & $1477.66$ & $-$ & $1493.81$ & $-$ \\
\hline

$w_0 w_a$\textbf{CDM} &&&&&& \\
BAO + CMB + SNeIa + $f\sigma_8$
& $-723.29$ & $1446.58$ & $0.889$ & $1456.58$ & $-12.10$ & $1483.57$ & $-1.30$ \\

BAO + CMB + SNeIa + $f+\sigma_8$
& $-721.03$ & $1442.06$ & $0.899$ & $1452.06$ & $-10.40$ & $1478.97$ & $0.36$ \\

BAO + CMB + SNeIa + $f\sigma_8 + f+\sigma_8$
& $-728.57$ & $1457.14$ & $0.91$ & $1467.14$ & $-10.52$ & $1494.06$ & $0.25$ \\
\hline

\textbf{Hu \& Sawicki} &&&&&& \\
BAO + CMB + SNeIa + $f\sigma_8$
& $-728.44$ & $1456.87$ & $0.894$ & $1465.87$ & $-2.81$ & $1486.47$ & $1.6$ \\

BAO + CMB + SNeIa + $f+\sigma_8$
& $-725.21$ & $1450.43$ & $0.904$ & $1458.43$ & $-4.03$ & $1479.96$ & $1.35$ \\

BAO + CMB + SNeIa + $f\sigma_8 + f+\sigma_8$
& $-732.78$ & $1465.55$ & $0.913$ & $1473.55$ & $-4.11$ & $1495.09$ & $1.28$ \\
\hline
\textbf{Starobinsky} &&&&&& \\
BAO + CMB + SNeIa + $f\sigma_8$ &  $-728.35$ & $1456.69$ & $0.894$ & $1464.69$ & $-3.99$ & $1486.29$ & $1.42$ \\

BAO + CMB + SNeIa + $f$ + $\sigma_8$ & $-725.01$ & $1450.01$ & $0.903$ & $1458.01$ & $-4.45$ & $1479.55$ & $0.94$ \\

BAO + CMB + SNeIa + $f\sigma_8$ + $f$ + $\sigma_8$ & $-732.59$ & $1465.18$ & $0.913$ & $1473.18$ & $-4.48$ & $1494.71$ & $0.90$  \\
\hline

\end{tabular}
}

\caption{
The values of $\mathcal{L(\hat{\theta},}D),\, \chi^2,\, \chi^2-\text{reduced},\, \text{AIC},\, \Delta\text{AIC},\, \text{BIC},\, \Delta\text{BIC}$ for $w_0w_a$CDM,\, and Hu \& Sawicki and Starobinsky models with $\Lambda$CDM taken as a benchmark model. According to the Jeffrey scale interpretation, the aforementioned models appear to be preferred over $\Lambda$CDM for the considered datasets (i.e., $\Delta\text{IC}_{\text{A,B}}<0$ for all models).
}
\label{tab2: comparison}

\end{table*}


\section{Constructing 2D phase-space diagrams using the LSS observables}\label{sec: LSS functions}
Deviations from $\Lambda$CDM are often described using phenomenological parameterizations for the LSS functions which include $\gamma$ and $\Sigma$ (see \cite{DESI:2024hhd,pogosian2010optimally, Ferte:2017bpf}). However, the authors in Ref.~\cite{Perenon:2015sla} demonstrated that several of these phenomenological parameterizations fail to reproduce some of the characteristic behaviors predicted by the full Effective Field Theory (EFT) of DE for modified gravity models. Although the latter article considered Horndeski models, our work studies both the time (or redshift) and scale dependence of the functions $\{\mu, \gamma, \Sigma\}$ in the context of the Hu \& Sawicki and Starobinsky $f(R)$ models. 

In fact, in~\cite{Perenon:2016blf}, the authors suggested a diagnostic in which Horndeski theories could be ruled out using the 2D planes, $\mu - \Sigma$ and $f\sigma_8-\sigma_8$. Thus, if future observations with model-independent predictions of $\Sigma - 1 < 1$ at $z = 0$ or $\mu -1 < 1$ with $\Sigma -1 > 1$ at $z > 1$, then Horndeski theories could effectively be ruled out. In the current work, we will make an analogous diagnostic for both the Hu \& Sawicki and Starobinsky models. Although this is a $f(R)$ model-dependent approach, given the generality of the two paradigmatic $f(R)$ models under consideration, it will serve as a null hypothesis for Hu \& Sawicki and Starobinsky $f(R)$ models.

\begin{figure*}
\centering
\includegraphics[width = 0.8\textwidth]{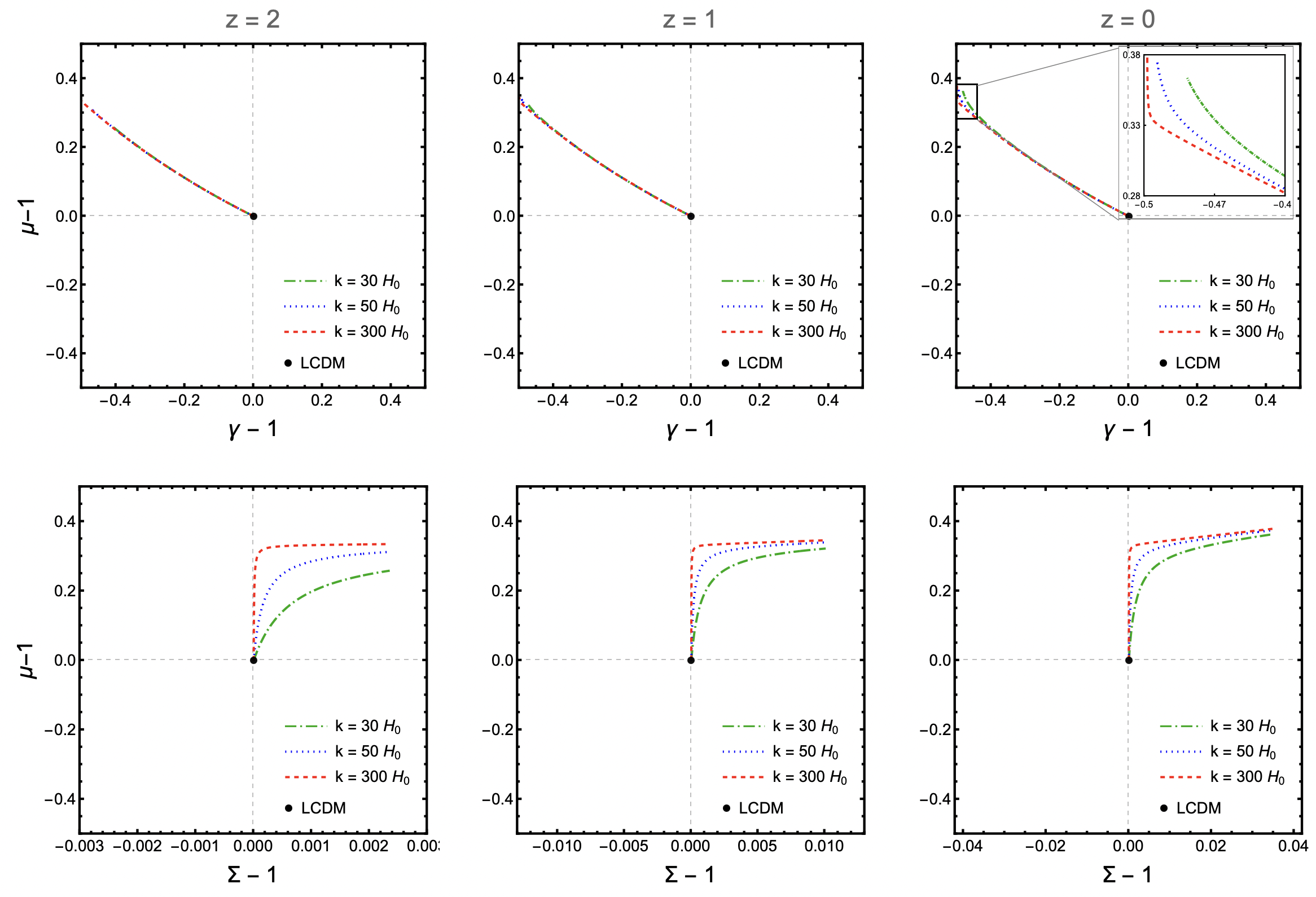}
\caption{The scale and redshift evolution of the LSS functions $\{\mu, \Sigma, \gamma\}$ for the Hu \& Sawicki $f(R)$ model. The dot represents the $\Lambda$CDM model where $\mu -1, \gamma-1, \text{ and } \Sigma-1$ are zero. The redshits are $z = 2, 1, 0$ while we only considered scales in the linear regime $k = 300 H_0,\, 50H_0,\, 30H_0$ represented by the dashed red, dotted blue, and dash-dotted green curves, respectively.}
\label{fig:H&S_LSS_funtion}
\end{figure*}

\begin{figure*}
\centering
\includegraphics[width = 0.8\textwidth]{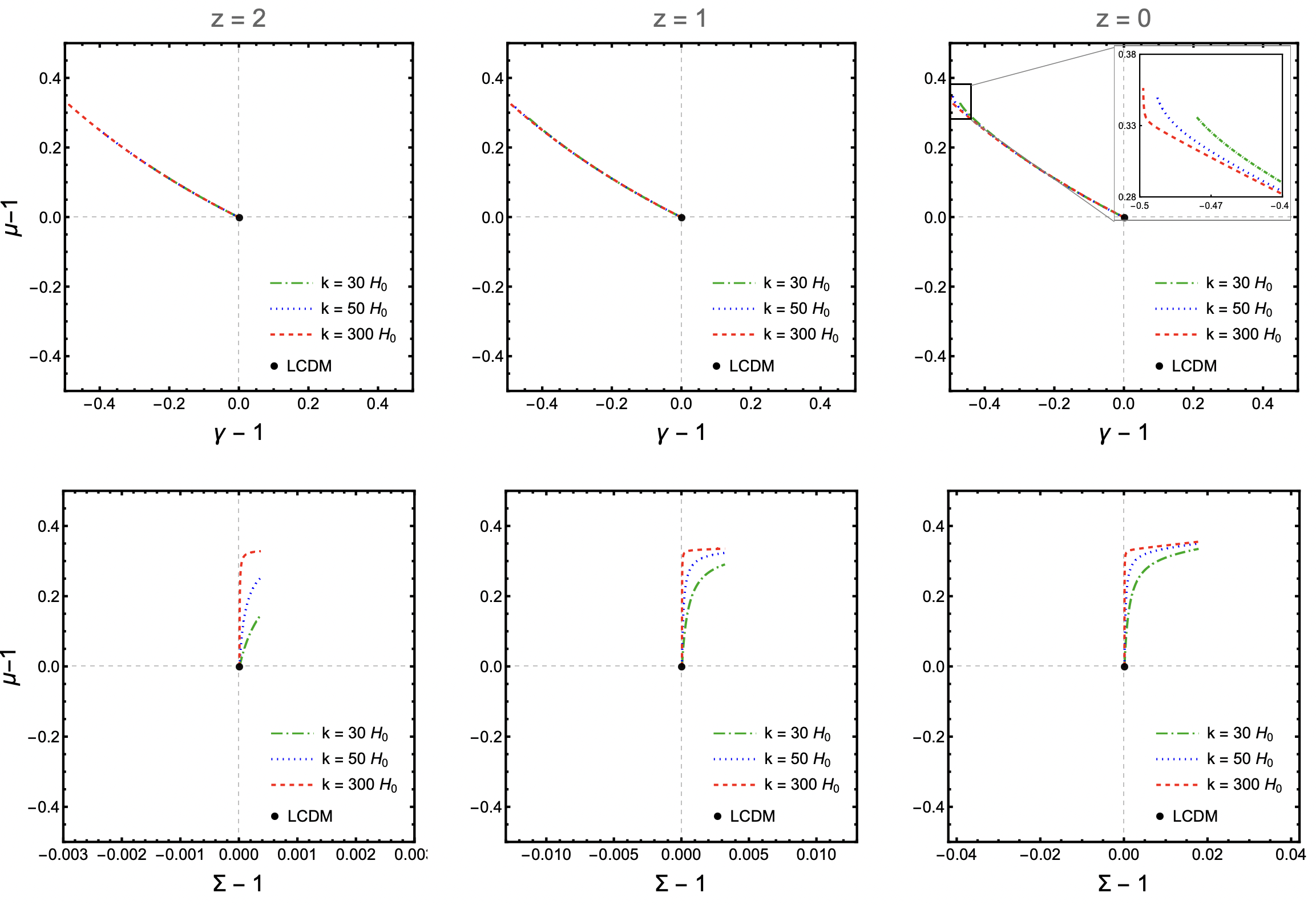}
\caption{The scale and redshift evolution of the LSS functions $\{\mu, \Sigma, \gamma\}$ for the Starobinsky $f(R)$ model. The dot represents the $\Lambda$CDM model where $\mu -1, \gamma-1, \text{ and } \Sigma-1$ are zero. The redshifts are $z = 2, 1, 0$ while we only considered scales in the linear regime $k = 300 H_0,\, 50H_0,\, 30H_0$ represented by the dashed red, dotted blue, and dash-dotted green curves, respectively.}
\label{fig:Starobinsky_LSS_funtion}
\end{figure*}

In Figs. \ref{fig:H&S_LSS_funtion} and \ref{fig:Starobinsky_LSS_funtion} we show the results for the evolution of the LSS observables, namely $\mu-1$ vs. $\gamma-1$ and $\mu-1$ vs. $\Sigma-1$, both in redshift and wavenumber, i.e., scale. In particular, for the Hu \& Sawicki model we have considered a range of values $b_{\rm HS}\in \{10^{-1.5}, 10^{-7.5}\}$, whereas $b_{\rm S} \in \{0.001, 0.6\}$ for the Starobinsky model.  These chosen intervals enclose values of $\text{log}_{10}b_{\rm HS}$ and $b_{\rm S}$ within $3\sigma$ bounds as shown in Figs. \ref{fig:H&S_MCMC} and \ref{fig:Starobinsky_MCMC}. We perform this procedure for the scales $30H_0$, $50H_0$, and $300H_0$. These scales all lie within the linear regime and therefore satisfy the sub-horizon approximation adopted for the evolution of density perturbations since Sec. \ref{sec: Perturbations}. The $30H_0$ and $50H_0$ scales were chosen for visual purposes, as the differences between the curves are clearly seen.

\section{Conclusions}\label{sec: conclusion}

In this work, we used the latest large-scale structure datasets to perform MCMC analyses to constrain the parameters for two paradigmatic $f(R)$ gravity models capable of providing late-time cosmic acceleration, namely the Hu \& Sawicki and Starobinsky models, and compared their performance with both $\Lambda$CDM and $w_0w_a$CDM. The resulting posterior distributions are shown as confidence contours in Figs. \ref{fig:LCDM MCMC}, \ref{fig:w0waCDM_MCMC}, \ref{fig:H&S_MCMC}, and \ref{fig:Starobinsky_MCMC}, for  $\Lambda$CDM and $w_0w_a$CDM and Hu \& Sawicki and Starobinsky models, respectively. In particular, Fig.~\ref{fig:H&S_MCMC} and Fig.~\ref{fig:Starobinsky_MCMC} show closed contours for $\Omega_\mathrm{m,0}, \sigma_{8,0}, h \text{ and the Starobinsky } b_{\rm S}$ parameter, while the Hu \& Sawicki parameter $\text{log}_{10}b_{\rm HS}$ is only bounded from above. The best-fit values with their corresponding 1$\sigma$ errors can be found in Table~\ref{tab1: bestfitparameters}. 

Using the AIC and BIC information criteria we presented the values of $\Delta$AIC and $\Delta$BIC using the $\Lambda$CDM as the base model. We noted that for all the three competing models under study, the values of $\Delta$AIC are negative implying that for the dataset combinations utilized in this communication, the alternative models are preferred. This is seen more strongly for the $w_0w_a$CDM possessing $|\Delta\text{AIC}| > 10$. For both the Hu \& Sawicki and Starobinsky models we noted that they lie between $4\le |\Delta\text{AIC}|\le7$ and according to the Jeffrey's scale, this implies a mild support for such models compared to the $\Lambda$CDM model for the considered datasets. The discrepancy is seen for Hu \& Sawicki model where the combined dataset $\text{BAO} + \text{CMB} + \text{SNeIa} + f\sigma_8$ suggests that $\Delta\text{AIC}<4$. This implies that for this dataset combination, the model is indistinguishable from the $\Lambda$CDM model.
However, for all the considered models, the evidence for support is less compelling when model complexity is taken into account since $0\le|\Delta\text{BIC}|\le2$.

Given the importance of weak lensing surveys in constraining modified gravity theories through the LSS functions, which include $\{\gamma, \Sigma\}$, we provided the analysis for those functions in the context of $f(R)$ gravity models. The evolution was studied both in the time (or redshift) and scale evolution for the Hu \& Sawicki and Starobinsky models. The results are shown in Figs. \ref{fig: Gamma&Sigma_H&S} and \ref{fig: Gamma&Sigma_Starobinsky}, respectively. In particular, we vary the parameters $b_{\rm HS}$ and $b_{\rm S}$ for the Hu \& Sawicki and Starobinsky models between the values $10^{-1} - 10^{-6}$ for $b$ and $0.001 -0.6$ for $\beta$. 
At early-times, $\Sigma = 1$ and $\gamma = 1$, i.e they both follow the $\Lambda$CDM evolution, while at late-times, they start to deviate from this scenario. More specifically, the deviations from $\Lambda$CDM for the Hu \& Sawicki model occur at $z \sim 19$ for the $\Sigma$ and at $z \sim 4.5$ for $\gamma$. While for the Starobinsky model, the deviations occur at $z \sim 8.5$ for $\Sigma$ and at $z\sim 3.1 $ for $\gamma$.

Another important feature is that the quantity $\gamma$ approaches a maximum value of $1/2$ at late times. This behaviour can be understood within the sub-Horizon approximation and by considering the late-time regime in which $\frac{k^2}{a^2}\frac{f_{RR}}{f_R}\gg1$ \cite{Tsujikawa:2007gd}. In this regime, the terms proportional to $\frac{k^2}{a^2}\frac{f_{RR}}{f_R}$ dominate in Eq.~(\ref{eq: gamma}), yielding $\gamma \rightarrow \frac{1}{2}$. Conversely, in the regime $\frac{k^2}{a^2}\frac{f_{RR}}{f_R}\ll1,$ the scale-dependent contributions become negligible, and Eq.~(\ref{eq: gamma}) reduces to $\gamma\rightarrow1$. Thus, the latter limit corresponds to the GR regime, whereas the $\gamma\rightarrow1/2$ limit characterizes the modified-gravity regime in which the additional scalar degree of freedom associated with $f(R)$ becomes dynamically relevant.

We also provide the scale evolution of the quantities $\{\mu, \gamma, \Sigma\}$. We produce a 2D ($\mu, \gamma$) and ($\mu, \Sigma$) phase-space diagrams
in Fig. \ref{fig:H&S_LSS_funtion} for the Hu \& Sawicki model and in Fig. \ref{fig:Starobinsky_LSS_funtion} for the Starobinsky model. In ($\mu, \Sigma$) phase-space diagram, $\mu$ seem to plateau at a value of approximately 1.38, a feature that is seen in both models for $z = 0$. The latter result can be clearly seen when one considers the regime (i) in Eq.~(\ref{eq: Geff&Qeff}). In the same phase-space, $\Sigma$ reaches a maximum at different values across all redshifts. In particular, at $z = 0$, $\Sigma \sim 1.038$, and $\Sigma \sim 1.018$ for the Hu \& Sawicki and Starobinsky models respectively. Although it is clear that the curves in the ($\mu, \Sigma$) diagrams are not equal for the depicted scales, there seem to occupy the first quadrant (i.e., when $\mu > 0$ and $\Sigma>0$). While in ($\mu, \gamma$) diagram, they occupy the second quadrant (i.e. when $\mu >0$ and $\gamma <0$. Those features are seen for both models across all the considered redshifts. Hence, if future surveys suggeste values of $\mu, \gamma, \Sigma$ not located in the aforementioned quadrants, the Hu \& Sawicki and Starobinsky models would be effectively ruled out by using these observables. 

\section*{Acknowledgements}
TM is grateful to the Instituto de Física Teórica (UAM-CSIC, Spain) for its warm hospitality during his internship there, when a substantial part of this work took place. TM acknowledges financial support from National Astrophysics and Space Science Programme (NASSP, South Africa) and the Alliance4Universities through its Erasmus+ KA171 programme. SN acknowledges support from the research project PID2024-159420NB-C43, the Proyecto de Investigaci\'on SAFE25003 from the Consejo Superior de Investigaciones Cient\'ificas (CSIC). This publication has been funded within the framework of the R\&D\&I Project CEX2025-001574-S, funded by MICIU/AEI/10.13039/501100011033. The research presented in this publication falls within the research line Origin and Composition of the Universe: Astroparticles and Cosmology (Astro/Cosmo). %
AdlCD acknowledges support from Project SA097P24 funded by Junta de Castilla y Le\'on (Spain), PID2024-158938NBI00 and CNS2024-154286 funded by MCIN/AEI/10.13039/\- 501100011033 and by {\it ERDF A way of making Europe}, and  NRF Grant CSUR23042798041 (South Africa).\\

\section{Appendix}

\begin{table}
\small
\centering
\begin{tabular}{c c c c c}
\hline
$z$ & $f\sigma_8$ & $f$ & $\sigma_8$ & Ref. \\
\hline\hline
0.001 & 0.505 $\pm$ 0.085 & & & \cite{Howlett:2017asq}\\
0.02 & 0.428 $\pm$ 0.0465 & & & \cite{Huterer:2016uyq}\\
0.02 & 0.398 $\pm$ 0.065 & & & \cite{hudson2012growth,turnbull2012cosmic}\\ 
0.02 & 0.314 $\pm$ 0.048 & & & \cite{hudson2012growth,Davis:2010sw}\\ 
0.10 & 0.376 $\pm$ 0.038 & 0.464 $\pm$ 0.040 & 0.769 $\pm$ 0.105 & \cite{Shi:2017qpr}\\ 
0.15 & 0.490 $\pm$ 0.145 & & & \cite{Howlett:2014opa}\\ 
0.17 & 0.510 $\pm$ 0.060 & & & \cite{Song:2008qt}\\ 
0.18 & 0.360 $\pm$ 0.090 & & & \cite{Blake:2013nif}\\ 
0.38 & 0.440 $\pm$ 0.060 & & &\cite{Blake:2013nif}\\ 
0.25 & 0.3512 $\pm$ 0.0583 & & & \cite{samushia2012interpreting}\\ 
0.37 & 0.4602 $\pm$ 0.0378 & & & \cite{samushia2012interpreting}\\ 
0.31 & 0.469 $\pm$ 0.098 & & &  \cite{Wang:2017wia}\\
0.36 & 0.474 $\pm$ 0.097 & & &  \cite{Wang:2017wia}\\
0.40 & 0.473 $\pm$ 0.086 & & &  \cite{Wang:2017wia}\\
0.44 & 0.481 $\pm$ 0.076 & & &  \cite{Wang:2017wia}\\
0.48 & 0.482 $\pm$ 0.067 & & &  \cite{Wang:2017wia}\\
0.52 & 0.488 $\pm$ 0.065 & & &  \cite{Wang:2017wia}\\
0.56 & 0.482 $\pm$ 0.067 & & &  \cite{Wang:2017wia}\\
0.59  & 0.481 $\pm$ 0.066 & & &  \cite{Wang:2017wia}\\
0.64  & 0.486 $\pm$ 0.070 & & &  \cite{Wang:2017wia}\\
0.44  & 0.413 $\pm$ 0.080 & & & \cite{blake2012wigglez}\\ 
0.60  & 0.390 $\pm$ 0.063 & & & \cite{blake2012wigglez}\\ 
0.73  & 0.437 $\pm$ 0.072 & & & \cite{blake2012wigglez}\\ 
0.60  & 0.550 $\pm$ 0.120 & 0.93 $\pm$ 0.22 & 0.52 $\pm$ 0.06 & \cite{delaTorre:2016rxm,Pezzotta:2016gbo}\\ 
0.86  & 0.400 $\pm$ 0.110 & 0.99 $\pm$ 0.19 & 0.48 $\pm$ 0.04 & \cite{delaTorre:2016rxm,Pezzotta:2016gbo}\\ 
1.40  & 0.482 $\pm$ 0.116 & & & \cite{Okumura:2015lvp}\\ 
0.978 & 0.379 $\pm$ 0.176 & & &  \cite{eBOSS:2018yfg}\\
1.23  & 0.385 $\pm$ 0.099 & & &  \cite{eBOSS:2018yfg}\\
1.526 & 0.342 $\pm$ 0.070 & & &  \cite{eBOSS:2018yfg}\\
1.944 & 0.364 $\pm$ 0.106 & & &  \cite{eBOSS:2018yfg}\\
\hline
\end{tabular} 
\caption{Compilation of growth of structure data obtained from various RSD surveys. The three data points for $f$ and $\sigma_8$ as obtained from \cite{Shi:2017qpr} and \cite{Pezzotta:2016gbo} are also shown.}
\label{tab:fsigma_f_sigma_data}
\end{table}

In  Table \ref{tab:fsigma_f_sigma_data} we show an updated compilation of $f\sigma_8$, $f$, and $\sigma_8$ data points as reported in the corresponding references (last column of  Table \ref{tab:fsigma_f_sigma_data}).

\bibliography{bibliograph}

@article{Abdelwahab:2011dk,
    author = "Abdelwahab, Mohamed and Goswami, Rituparno and Dunsby, Peter K. S.",
    title = "{Cosmological dynamics of fourth order gravity: A compact view}",
    eprint = "1111.0171",
    archivePrefix = "arXiv",
    primaryClass = "gr-qc",
    doi = "10.1103/PhysRevD.85.083511",
    journal = "Phys. Rev. D",
    volume = "85",
    pages = "083511",
    year = "2012"
}

@article{Carloni_2007,
   title={A dynamical system approach to higher order gravity},
   volume={40},
   ISSN={1751-8121},
   url={http://dx.doi.org/10.1088/1751-8113/40/25/S40},
   DOI={10.1088/1751-8113/40/25/s40},
   number={25},
   journal={Journal of Physics A: Mathematical and Theoretical},
   publisher={IOP Publishing},
   author={Carloni, Sante and Dunsby, Peter K S},
   year={2007},
   month=jun, pages={6919–6925}
}

@article{Abdelwahab_2012,
   title={Cosmological dynamics of fourth-order gravity: A compact view},
   volume={85},
   ISSN={1550-2368},
   url={http://dx.doi.org/10.1103/PhysRevD.85.083511},
   DOI={10.1103/physrevd.85.083511},
   number={8},
   journal={Physical Review D},
   publisher={American Physical Society (APS)},
   author={Abdelwahab, Mohamed and Goswami, Rituparno and Dunsby, Peter K. S.},
   year={2012},
   month=apr 
   }

@article{de_la_Cruz_Dombriz_2016,
   title={Theoretical and observational constraints of viable theories of gravity},
   volume={93},
   ISSN={2470-0029},
   url={http://dx.doi.org/10.1103/PhysRevD.93.084016},
   DOI={10.1103/physrevd.93.084016},
   number={8},
   journal={Physical Review D},
   publisher={American Physical Society (APS)},
   author={de la Cruz-Dombriz, Alvaro and Dunsby, Peter K.S. and Kandhai, Sulona and Saez-Gomez, Diego},
   year={2016},
   month=apr }

@article{Planck18,
   title={Planck2018 results: VI. Cosmological parameters},
   volume={641},
   ISSN={1432-0746},
   url={http://dx.doi.org/10.1051/0004-6361/201833910},
   DOI={10.1051/0004-6361/201833910},
   journal={Astronomy \&; Astrophysics},
   publisher={EDP Sciences},
   author={Aghanim, N. et al},
   year={2020},
   month=sep, pages={A6} }

@article{Hu:2007nk,
    author = "Hu, Wayne and Sawicki, Ignacy",
    title = "{Models of f(R) Cosmic Acceleration that Evade Solar-System Tests}",
    eprint = "0705.1158",
    archivePrefix = "arXiv",
    primaryClass = "astro-ph",
    doi = "10.1103/PhysRevD.76.064004",
    journal = "Phys. Rev. D",
    volume = "76",
    pages = "064004",
    year = "2007"
}

@article{Nesseris:2012cq,
    author = "Nesseris, Savvas and Garcia-Bellido, Juan",
    title = "{Is the Jeffreys' scale a reliable tool for Bayesian model comparison in cosmology?}",
    eprint = "1210.7652",
    archivePrefix = "arXiv",
    primaryClass = "astro-ph.CO",
    reportNumber = "IFT-UAM-CSIC-12-95",
    doi = "10.1088/1475-7516/2013/08/036",
    journal = "JCAP",
    volume = "8",
    pages = "36",
    year = "2013"
}

@ARTICLE{Akaike,
  author={Akaike, H.},
  journal={IEEE Transactions on Automatic Control}, 
  title={A new look at the statistical model identification}, 
  year={1974},
  volume={19},
  number={6},
  pages={716-723},
  doi={10.1109/TAC.1974.1100705}
  }

@ARTICLE{Schwarz,
  author={Schwarz, G.},
  journal={The Annals of Statistics}, 
  title={Estimating the Dimension of a Model.}, 
  year={1978},
  volume={6},
  number={2},
  pages={461-464},
  doi={http://www.jstor.org/stable/2958889}
  }

@article{DESI:2025zgx,
    author = "Abdul Karim, M. and others",
    collaboration = "DESI",
    title = "{DESI DR2 results. II. Measurements of baryon acoustic oscillations and cosmological constraints}",
    eprint = "2503.14738",
    archivePrefix = "arXiv",
    primaryClass = "astro-ph.CO",
    reportNumber = "FERMILAB-PUB-25-0169-PPD",
    doi = "10.1103/tr6y-kpc6",
    journal = "Phys. Rev. D",
    volume = "112",
    number = "8",
    pages = "083515",
    year = "2025"
}

@article{Arjona:2018jhh,
    author = "Arjona, Rub{\'e}n and Cardona, Wilmar and Nesseris, Savvas",
    title = "{Unraveling the effective fluid approach for $f(R)$ models in the subhorizon approximation}",
    eprint = "1811.02469",
    archivePrefix = "arXiv",
    primaryClass = "astro-ph.CO",
    reportNumber = "IFT-UAM/CSIC-18-108",
    doi = "10.1103/PhysRevD.99.043516",
    journal = "Phys. Rev. D",
    volume = "99",
    number = "4",
    pages = "043516",
    year = "2019"
}

@article{Planck:2018vyg,
    author = "Aghanim, N. and others",
    collaboration = "Planck",
    title = "{Planck 2018 results. VI. Cosmological parameters}",
    eprint = "1807.06209",
    archivePrefix = "arXiv",
    primaryClass = "astro-ph.CO",
    doi = "10.1051/0004-6361/201833910",
    journal = "Astron. Astrophys.",
    volume = "641",
    pages = "A6",
    year = "2020",
    note = "[Erratum: Astron.Astrophys. 652, C4 (2021)]"
}

@article{Aizpuru:2021vhd,
    author = "Aizpuru, Andoni and Arjona, Rub{\'e}n and Nesseris, Savvas",
    title = "{Machine learning improved fits of the sound horizon at the baryon drag epoch}",
    eprint = "2106.00428",
    archivePrefix = "arXiv",
    primaryClass = "astro-ph.CO",
    reportNumber = "IFT-UAM/CSIC-21-67",
    doi = "10.1103/PhysRevD.104.043521",
    journal = "Phys. Rev. D",
    volume = "104",
    number = "4",
    pages = "043521",
    year = "2021"
}

@article{Euclid:2025bxg,
    author = "Ocampo, I. and others",
    collaboration = "Euclid",
    title = "{Euclid: Forecasts on $\Lambda$CDM consistency tests with growth rate data}",
    eprint = "2507.22780",
    archivePrefix = "arXiv",
    primaryClass = "astro-ph.CO",
    month = "7",
    journal = "{}",
    year = "2025"
}

@article{Song:2008qt,
    author = "Song, Yong-Seon and Percival, Will J.",
    title = "{Reconstructing the history of structure formation using Redshift Distortions}",
    eprint = "0807.0810",
    archivePrefix = "arXiv",
    primaryClass = "astro-ph",
    doi = "10.1088/1475-7516/2009/10/004",
    journal = "JCAP",
    volume = "10",
    pages = "004",
    year = "2009"
}

@article{DeFelice:2010aj,
    author = "De Felice, Antonio and Tsujikawa, Shinji",
    title = "{f(R) theories}",
    eprint = "1002.4928",
    archivePrefix = "arXiv",
    primaryClass = "gr-qc",
    doi = "10.12942/lrr-2010-3",
    journal = "Living Rev. Rel.",
    volume = "13",
    pages = "3",
    year = "2010"
}

@article{Kumar:2023bqj,
    author = "Kumar, Suresh and Nunes, Rafael C. and Pan, Supriya and Yadav, Priya",
    title = "{New late-time constraints on f(R) gravity}",
    eprint = "2301.07897",
    archivePrefix = "arXiv",
    primaryClass = "astro-ph.CO",
    doi = "10.1016/j.dark.2023.101281",
    journal = "Phys. Dark Univ.",
    volume = "42",
    pages = "101281",
    year = "2023"
}

@article{Starobinsky:2007hu,
    author = "Starobinsky, Alexei A.",
    title = "{Disappearing cosmological constant in f(R) gravity}",
    eprint = "0706.2041",
    archivePrefix = "arXiv",
    primaryClass = "astro-ph",
    doi = "10.1134/S0021364007150027",
    journal = "JETP Lett.",
    volume = "86",
    pages = "157--163",
    year = "2007"
}

@article{Tsujikawa:2007gd,
    author = "Tsujikawa, Shinji",
    title = "{Matter density perturbations and effective gravitational constant in modified gravity models of dark energy}",
    eprint = "0705.1032",
    archivePrefix = "arXiv",
    primaryClass = "astro-ph",
    doi = "10.1103/PhysRevD.76.023514",
    journal = "Phys. Rev. D",
    volume = "76",
    pages = "023514",
    year = "2007"
}

@phdthesis{delaCruzDombriz:2010xy,
    author = "de la Cruz Dombriz, Alvaro",
    title = "{Some cosmological and astrophysical aspects of modified gravity theories}",
    eprint = "1004.5052",
    archivePrefix = "arXiv",
    primaryClass = "gr-qc",
    school = "Madrid U.",
    year = "2010"
}

@article{Murakami:2023qdl,
    author = "Murakami, Koya and Ocampo, Indira and Nesseris, Savvas and Nishizawa, Atsushi J. and Kuroyanagi, Sachiko",
    title = "{Nonlinearity-free prediction of the growth-rate f{\ensuremath{\sigma}}8 using convolutional neural networks}",
    eprint = "2305.12812",
    archivePrefix = "arXiv",
    primaryClass = "astro-ph.CO",
    reportNumber = "IFT-UAM/CSIC-23-57",
    doi = "10.1103/PhysRevD.110.023525",
    journal = "Phys. Rev. D",
    volume = "110",
    number = "2",
    pages = "023525",
    year = "2024"
}

@article{Orjuela-Quintana:2023zjm,
    author = "Orjuela-Quintana, J. Bayron and Nesseris, Savvas",
    title = "{Tracking the validity of the quasi-static and sub-horizon approximations in modified gravity}",
    eprint = "2303.14251",
    archivePrefix = "arXiv",
    primaryClass = "gr-qc",
    reportNumber = "IFT-UAM/CSIC-23-27",
    doi = "10.1088/1475-7516/2023/08/019",
    journal = "JCAP",
    volume = "08",
    pages = "019",
    year = "2023"
}

@article{Nesseris:2017vor,
    author = "Nesseris, Savvas and Pantazis, George and Perivolaropoulos, Leandros",
    title = "{Tension and constraints on modified gravity parametrizations of $G_{\textrm{eff}}(z)$ from growth rate and Planck data}",
    eprint = "1703.10538",
    archivePrefix = "arXiv",
    primaryClass = "astro-ph.CO",
    reportNumber = "IFT-UAM-CSIC-17-031",
    doi = "10.1103/PhysRevD.96.023542",
    journal = "Phys. Rev. D",
    volume = "96",
    number = "2",
    pages = "023542",
    year = "2017"
}

@article{Perenon:2019dpc,
    author = "Perenon, Louis and Bel, Julien and Maartens, Roy and de la Cruz-Dombriz, Alvaro",
    title = "{Optimising growth of structure constraints on modified gravity}",
    eprint = "1901.11063",
    archivePrefix = "arXiv",
    primaryClass = "astro-ph.CO",
    doi = "10.1088/1475-7516/2019/06/020",
    journal = "JCAP",
    volume = "06",
    pages = "020",
    year = "2019"
}

@article{Howlett:2017asq,
    author = {Howlett, Cullan and Staveley-Smith, Lister and Elahi, Pascal J. and Hong, Tao and Jarrett, Tom H. and Jones, D. Heath and Koribalski, B{\"a}rbel S. and Macri, Lucas M. and Masters, Karen L. and Springob, Christopher M.},
    title = "{2MTF {\textendash} VI. Measuring the velocity power spectrum}",
    eprint = "1706.05130",
    archivePrefix = "arXiv",
    primaryClass = "astro-ph.CO",
    doi = "10.1093/mnras/stx1521",
    journal = "Mon. Not. Roy. Astron. Soc.",
    volume = "471",
    number = "3",
    pages = "3135--3151",
    year = "2017"
}

@article{Huterer:2016uyq,
    author = "Huterer, Dragan and Shafer, Daniel and Scolnic, Daniel and Schmidt, Fabian",
    title = "{Testing $\Lambda$CDM at the lowest redshifts with SN Ia and galaxy velocities}",
    eprint = "1611.09862",
    archivePrefix = "arXiv",
    primaryClass = "astro-ph.CO",
    doi = "10.1088/1475-7516/2017/05/015",
    journal = "JCAP",
    volume = "05",
    pages = "015",
    year = "2017"
}

@article{hudson2012growth,
  title={The growth rate of cosmic structure from peculiar velocities at low and high redshifts},
  author={Hudson, Michael J and Turnbull, Stephen J},
  journal={The Astrophysical Journal Letters},
  volume={751},
  number={2},
  pages={L30},
  year={2012},
  publisher={The American Astronomical Society}
}

@article{turnbull2012cosmic,
  title={Cosmic flows in the nearby universe from Type Ia Supernovae},
  author={Turnbull, Stephen J and Hudson, Michael J and Feldman, Hume A and Hicken, Malcolm and Kirshner, Robert P and Watkins, Richard},
  journal={Monthly Notices of the Royal Astronomical Society},
  volume={420},
  number={1},
  pages={447--454},
  year={2012},
  publisher={The Royal Astronomical Society}
}

@article{Davis:2010sw,
    author = "Davis, Marc and Nusser, Adi and Masters, Karen and Springob, Christopher and Huchra, John P. and Lemson, Gerard",
    title = "{Local Gravity versus Local Velocity: Solutions for $\beta$ and nonlinear bias}",
    eprint = "1011.3114",
    archivePrefix = "arXiv",
    primaryClass = "astro-ph.CO",
    doi = "10.1111/j.1365-2966.2011.18362.x",
    journal = "Mon. Not. Roy. Astron. Soc.",
    volume = "413",
    pages = "2906",
    year = "2011"
}

@article{Shi:2017qpr,
    author = "Shi, Feng and others",
    title = "{Mapping the Real Space Distributions of Galaxies in SDSS DR7: II. Measuring the growth rate, clustering amplitude of matter and biases of galaxies at redshift $0.1$}",
    eprint = "1712.04163",
    archivePrefix = "arXiv",
    primaryClass = "astro-ph.CO",
    doi = "10.3847/1538-4357/aacb20",
    journal = "Astrophys. J.",
    volume = "861",
    number = "2",
    pages = "137",
    year = "2018"
}

@article{Howlett:2014opa,
    author = "Howlett, Cullan and Ross, Ashley and Samushia, Lado and Percival, Will and Manera, Marc",
    title = "{The clustering of the SDSS main galaxy sample {\textendash} II. Mock galaxy catalogues and a measurement of the growth of structure from redshift space distortions at $z = 0.15$}",
    eprint = "1409.3238",
    archivePrefix = "arXiv",
    primaryClass = "astro-ph.CO",
    doi = "10.1093/mnras/stu2693",
    journal = "Mon. Not. Roy. Astron. Soc.",
    volume = "449",
    number = "1",
    pages = "848--866",
    year = "2015"
}

@article{Blake:2013nif,
    author = "Blake, Chris and others",
    title = "{Galaxy And Mass Assembly (GAMA): improved cosmic growth measurements using multiple tracers of large-scale structure}",
    eprint = "1309.5556",
    archivePrefix = "arXiv",
    primaryClass = "astro-ph.CO",
    doi = "10.1093/mnras/stt1791",
    journal = "Mon. Not. Roy. Astron. Soc.",
    volume = "436",
    pages = "3089",
    year = "2013"
}

@article{samushia2012interpreting,
  title={Interpreting large-scale redshift-space distortion measurements},
  author={Samushia, Lado and Percival, Will J and Raccanelli, Alvise},
  journal={Monthly Notices of the Royal Astronomical Society},
  volume={420},
  number={3},
  pages={2102--2119},
  year={2012},
  publisher={Blackwell Publishing Ltd Oxford, UK}
}

@article{Wang:2017wia,
    author = "Wang, Yuting and Zhao, Gong-Bo and Chuang, Chia-Hsun and Pellejero-Ibanez, Marcos and Zhao, Cheng and Kitaura, Francisco-Shu and Rodriguez-Torres, Sergio",
    title = "{The clustering of galaxies in the completed SDSS-III Baryon Oscillation Spectroscopic Survey: a tomographic analysis of structure growth and expansion rate from anisotropic galaxy clustering}",
    eprint = "1709.05173",
    archivePrefix = "arXiv",
    primaryClass = "astro-ph.CO",
    doi = "10.1093/mnras/sty2449",
    journal = "Mon. Not. Roy. Astron. Soc.",
    volume = "481",
    number = "3",
    pages = "3160--3166",
    year = "2018"
}

@article{blake2012wigglez,
  title={The WiggleZ Dark Energy Survey: Joint measurements of the expansion and growth history at z< 1},
  author={Blake, Chris and Brough, Sarah and Colless, Matthew and Contreras, Carlos and Couch, Warrick and Croom, Scott and Croton, Darren and Davis, Tamara M and Drinkwater, Michael J and Forster, Karl and others},
  journal={Monthly Notices of the Royal Astronomical Society},
  volume={425},
  number={1},
  pages={405--414},
  year={2012},
  publisher={Blackwell Science Ltd Oxford, UK}
}

@article{delaTorre:2016rxm,
    author = "de la Torre, S. and others",
    title = "{The VIMOS Public Extragalactic Redshift Survey (VIPERS). Gravity test from the combination of redshift-space distortions and galaxy-galaxy lensing at $0.5 < z < 1.2$}",
    eprint = "1612.05647",
    archivePrefix = "arXiv",
    primaryClass = "astro-ph.CO",
    doi = "10.1051/0004-6361/201630276",
    journal = "Astron. Astrophys.",
    volume = "608",
    pages = "A44",
    year = "2017"
}

@article{Pezzotta:2016gbo,
    author = "Pezzotta, A. and others",
    title = "{The VIMOS Public Extragalactic Redshift Survey (VIPERS): The growth of structure at $0.5 < z < 1.2$ from redshift-space distortions in the clustering of the PDR-2 final sample}",
    eprint = "1612.05645",
    archivePrefix = "arXiv",
    primaryClass = "astro-ph.CO",
    doi = "10.1051/0004-6361/201630295",
    journal = "Astron. Astrophys.",
    volume = "604",
    pages = "A33",
    year = "2017"
}

@article{Okumura:2015lvp,
    author = "Okumura, Teppei and others",
    title = "{The Subaru FMOS galaxy redshift survey (FastSound). IV. New constraint on gravity theory from redshift space distortions at $z\sim 1.4$}",
    eprint = "1511.08083",
    archivePrefix = "arXiv",
    primaryClass = "astro-ph.CO",
    doi = "10.1093/pasj/psw029",
    journal = "Publ. Astron. Soc. Jap.",
    volume = "68",
    number = "3",
    pages = "38",
    year = "2016"
}

@article{eBOSS:2018yfg,
    author = "Zhao, Gong-Bo and others",
    collaboration = "eBOSS",
    title = "{The clustering of the SDSS-IV extended Baryon Oscillation Spectroscopic Survey DR14 quasar sample: a tomographic measurement of cosmic structure growth and expansion rate based on optimal redshift weights}",
    eprint = "1801.03043",
    archivePrefix = "arXiv",
    primaryClass = "astro-ph.CO",
    doi = "10.1093/mnras/sty2845",
    journal = "Mon. Not. Roy. Astron. Soc.",
    volume = "482",
    number = "3",
    pages = "3497--3513",
    year = "2019"
}

@article{DESI:2024hhd,
    author = "Adame, A. G. and others",
    collaboration = "DESI",
    title = "{DESI 2024 VII: cosmological constraints from the full-shape modeling of clustering measurements}",
    eprint = "2411.12022",
    archivePrefix = "arXiv",
    primaryClass = "astro-ph.CO",
    reportNumber = "FERMILAB-PUB-24-0854-PPD",
    doi = "10.1088/1475-7516/2025/07/028",
    journal = "JCAP",
    volume = "07",
    pages = "028",
    year = "2025"
}

@article{pogosian2010optimally,
  title={How to optimally parametrize deviations from general relativity in the evolution<? format?> of cosmological perturbations},
  author={Pogosian, Levon and Silvestri, Alessandra and Koyama, Kazuya and Zhao, Gong-Bo},
  journal={Physical Review D—Particles, Fields, Gravitation, and Cosmology},
  volume={81},
  number={10},
  pages={104023},
  year={2010},
  publisher={APS}
}

@article{Perenon:2015sla,
    author = "Perenon, Louis and Piazza, Federico and Marinoni, Christian and Hui, Lam",
    title = "{Phenomenology of dark energy: general features of large-scale perturbations}",
    eprint = "1506.03047",
    archivePrefix = "arXiv",
    primaryClass = "astro-ph.CO",
    doi = "10.1088/1475-7516/2015/11/029",
    journal = "JCAP",
    volume = "11",
    pages = "029",
    year = "2015"
}

@article{Khoury:2003aq,
    author = "Khoury, Justin and Weltman, Amanda",
    title = "{Chameleon fields: Awaiting surprises for tests of gravity in space}",
    eprint = "astro-ph/0309300",
    archivePrefix = "arXiv",
    doi = "10.1103/PhysRevLett.93.171104",
    journal = "Phys. Rev. Lett.",
    volume = "93",
    pages = "171104",
    year = "2004"
}

@article{Pan-STARRS1:2017jku,
    author = "Scolnic, D. M. and others",
    collaboration = "Pan-STARRS1",
    title = "{The Complete Light-curve Sample of Spectroscopically Confirmed SNe Ia from Pan-STARRS1 and Cosmological Constraints from the Combined Pantheon Sample}",
    eprint = "1710.00845",
    archivePrefix = "arXiv",
    primaryClass = "astro-ph.CO",
    doi = "10.3847/1538-4357/aab9bb",
    journal = "Astrophys. J.",
    volume = "859",
    number = "2",
    pages = "101",
    year = "2018"
}

@article{Ferte:2017bpf,
    author = "Fert{\'e}, Agn{\`e}s and Kirk, Donnacha and Liddle, Andrew R. and Zuntz, Joe",
    title = "{Testing gravity on cosmological scales with cosmic shear, cosmic microwave background anisotropies, and redshift-space distortions}",
    eprint = "1712.01846",
    archivePrefix = "arXiv",
    primaryClass = "astro-ph.CO",
    doi = "10.1103/PhysRevD.99.083512",
    journal = "Phys. Rev. D",
    volume = "99",
    number = "8",
    pages = "083512",
    year = "2019"
}

@article{Perenon:2016blf,
    author = "Perenon, Louis and Marinoni, Christian and Piazza, Federico",
    title = "{Diagnostic of Horndeski Theories}",
    eprint = "1609.09197",
    archivePrefix = "arXiv",
    primaryClass = "astro-ph.CO",
    doi = "10.1088/1475-7516/2017/01/035",
    journal = "JCAP",
    volume = "01",
    pages = "035",
    year = "2017"
}

@article{Padmanabhan:2002ji,
    author = "Padmanabhan, T.",
    title = "{Cosmological constant: The Weight of the vacuum}",
    eprint = "hep-th/0212290",
    archivePrefix = "arXiv",
    doi = "10.1016/S0370-1573(03)00120-0",
    journal = "Phys. Rept.",
    volume = "380",
    pages = "235--320",
    year = "2003"
}

@article{Du:2026cly,
    author = "Du, Guo-Hong and Li, Tian-Nuo and Liu, Tonghua and Zhang, Jing-Fei and Zhang, Xin",
    title = "{Evidence for deviation in gravitational light deflection from general relativity at cosmological scales with KiDS-Legacy and CMB lensing}",
    eprint = "2602.03110",
    archivePrefix = "arXiv",
    primaryClass = "astro-ph.CO",
    doi = "10.1007/s11433-026-2987-7",
    journal = "Sci. China Phys. Mech. Astron.",
    volume = "69",
    number = "8",
    pages = "280411",
    year = "2026"
}

@article{DES:2024jxu,
    author = "Abbott, T. M. C. and others",
    collaboration = "DES",
    title = "{The Dark Energy Survey: Cosmology Results with {\ensuremath{\sim}}1500 New High-redshift Type Ia Supernovae Using the Full 5 yr Data Set}",
    eprint = "2401.02929",
    archivePrefix = "arXiv",
    primaryClass = "astro-ph.CO",
    reportNumber = "FERMILAB-PUB-23-0821-PPD, DES-2023-805",
    doi = "10.3847/2041-8213/ad6f9f",
    journal = "Astrophys. J. Lett.",
    volume = "973",
    number = "1",
    pages = "L14",
    year = "2024"
}

@article{Lopez:2024kih,
    author = "Lopez, Alexia M. and Clowes, Roger G. and Williger, Gerard M.",
    title = "{Investigating ultra-large large-scale structures: potential implications for cosmology}",
    eprint = "2409.14894",
    archivePrefix = "arXiv",
    primaryClass = "astro-ph.CO",
    doi = "10.1098/rsta.2024.0029",
    journal = "Phil. Trans. Roy. Soc. Lond. A",
    volume = "383",
    number = "2290",
    pages = "20240029",
    year = "2025"
}

@article{Kumar:2025mzo,
    author = "Kumar, Darshan and Dhankar, Praveen Kumar and Ray, Saibal and Zhang, Fengge",
    title = "{Joint analysis of constraints on f(R) parametrization from recent cosmological observations}",
    eprint = "2504.04118",
    archivePrefix = "arXiv",
    primaryClass = "astro-ph.CO",
    doi = "10.1016/j.dark.2025.101989",
    journal = "Phys. Dark Univ.",
    volume = "49",
    pages = "101989",
    year = "2025"
}

@article{Capozziello:2007eu,
    author = "Capozziello, Salvatore and Tsujikawa, Shinji",
    title = "{Solar system and equivalence principle constraints on f(R) gravity by chameleon approach}",
    eprint = "0712.2268",
    archivePrefix = "arXiv",
    primaryClass = "gr-qc",
    doi = "10.1103/PhysRevD.77.107501",
    journal = "Phys. Rev. D",
    volume = "77",
    pages = "107501",
    year = "2008"
}

@article{delaCruz-Dombriz:2015tye,
    author = "de la Cruz-Dombriz, {\'A}lvaro and Dunsby, Peter K. S. and Kandhai, Sulona and S{\'a}ez-G{\'o}mez, Diego",
    title = "{Theoretical and observational constraints of viable f(R) theories of gravity}",
    eprint = "1511.00102",
    archivePrefix = "arXiv",
    primaryClass = "gr-qc",
    doi = "10.1103/PhysRevD.93.084016",
    journal = "Phys. Rev. D",
    volume = "93",
    number = "8",
    pages = "084016",
    year = "2016"
}

@article{Kandhai:2015pyr,
    author = "Kandhai, Sulona and Dunsby, Peter K. S.",
    title = "{Cosmological dynamics of viable f(R) theories of gravity}",
    eprint = "1511.00101",
    archivePrefix = "arXiv",
    primaryClass = "gr-qc",
    month = "10",
    year = "2015",
    journal = ""
}

@article{Chevallier:2000qy,
    author = "Chevallier, Michel and Polarski, David",
    title = "{Accelerating universes with scaling dark matter}",
    eprint = "gr-qc/0009008",
    archivePrefix = "arXiv",
    doi = "10.1142/S0218271801000822",
    journal = "Int. J. Mod. Phys. D",
    volume = "10",
    pages = "213--224",
    year = "2001"
}

@article{Linder:2002et,
    author = "Linder, Eric V.",
    title = "{Exploring the expansion history of the universe}",
    eprint = "astro-ph/0208512",
    archivePrefix = "arXiv",
    doi = "10.1103/PhysRevLett.90.091301",
    journal = "Phys. Rev. Lett.",
    volume = "90",
    pages = "091301",
    year = "2003"
}

@article{Multamaki:2005zs,
    author = "Multamaki, T. and Vilja, Iiro",
    title = "{Cosmological expansion and the uniqueness of gravitational action}",
    eprint = "astro-ph/0506692",
    archivePrefix = "arXiv",
    reportNumber = "NORDITA-2005-43",
    doi = "10.1103/PhysRevD.73.024018",
    journal = "Phys. Rev. D",
    volume = "73",
    pages = "024018",
    year = "2006"
}

\end{document}